\documentclass[preprint2]{aastex6}
\usepackage{epsfig}
\usepackage{amsfonts,amssymb,amsmath}

\begin{document}

\title{On the Mass Distribution of the Intra-Cluster Light in Galaxy Groups and Clusters}
\author{E. Contini$^{1,2}$ and Q. Gu$^{1,2}$}

\affil{$^1$School of Astronomy and Space Science, Nanjing University, Nanjing 210093, China; {\color{blue} emanuele.contini82@gmail.com, qsgu@nju.edu.cn}}
\affil{$^2$Key Laboratory of Modern Astronomy and Astrophysics (Nanjing University), Ministry of Education, China}

\email{emanuele.contini82@gmail.com}
\email{qsgu@nju.edu.cn}

\begin{abstract} 
We take advantage of a semi-analytic model with a state-of-art implementation of the formation of the intra-cluster light (ICL) to study the mass distribution of the ICL in galaxy 
groups and clusters, at different redshifts. We assume the ICL to follow a NFW profile with a different concentration, linked to that of the dark matter by the relation $c_{ICL}=\gamma c_{DM}$, 
where the parameter $\gamma$ is set to reproduce the observed relation between the stellar mass in the brightest cluster galaxy (BCG) and ICL in the innermost 100 kpc and the halo 
mass ($M^*_{100}-M_{500}$), at $z=0$. The model is then tested against several theoretical and observational results, from the present time to $z\sim1.5$.
Our analysis shows that the fraction of stellar mass in the BCG and ICL within the innermost 100 kpc is an increasing function of redshift, parameter $\gamma$, and a decreasing 
function of the halo mass. The value of $\gamma$ required to match the observed $M^*_{100}-M_{500}$ is $\gamma=3$ at $z=0$, but there are hints that it might be a function of 
redshift and halo mass. This result indicates that the distribution of the ICL is more concentrated than the dark matter one, but less concentrated than previously found by other studies.  
We suggest that a modified version of the NFW is a good description of the distribution of the diffuse light in groups and clusters, which makes the ICL a reliable tracer of the dark 
matter, in good agreement with recent observational findings.
\end{abstract}


\section[]{Introduction} 
\label{sec:intro}

The intracluster light (ICL), which was first discovered by \cite{zwicky37}, is a diffuse component in galaxy groups and clusters made of stars that are not bound to any galaxy. It is commonly associated with the 
diffuse envelope of the most massive galaxies that reside in the center of groups and clusters (e.g., \citealt{gonzalez13,kravtsov18}), although a modest fraction has been found (e.g. \citealt{presotto14}) and 
predicted (e.g., \citealt{contini18}) to be around intermediate/massive satellites, especially in the most massive clusters in the local universe. Since its discovery, the ICL has been studied with the idea that it 
could shed some light on the processes at play in the formation of the large structures such as galaxy clusters.  

In the last decade or so, many attempts, both from the observational and theoretical sides, have been made in order to understand the physical mechanisms that bring to the formation of the ICL
(\citealt{murante07,purcell07,puchwein10,rudick11,contini14,demaio15,burke15,iodice17,groenewald17,morishita17,tang18,montes18,demaio18,contini18, montes19,demaio20,iodice20} and many others). 
Most of this diffuse light is found around the brightest cluster galaxy (BCG) and so, as a natural consequence, the main mechanisms invoked to explain the formation of this component are those related to the 
formation and evolution of the BCGs, such as galaxy mergers and stellar stripping. From a theoretical point of view, the mutual role of these two processes would bring to different properties of both BCGs and 
ICL, such as colors and metallicity (a detailed discussion of this topic can be found in \citealt{contini18,contini19}). 

Taking advantage of the latest version of the original model (presented in \citealt{contini14}) for the formation of the ICL, in \cite{contini18} we focused on the growth of the BCG and the ICL. Among the most relevant 
results obtained, we showed that BCGs and ICL form and grow at different times and with different rates (overall), but the two components co-evolve after redshift $z\sim0.7$. Stellar stripping is the most important process 
in our model that leads to the formation of the ICL, and, around 90\% of the ICL coming from this channel is actually produced in the innermost 150 kpc from the halo center, with a significant halo-to-halo scatter that 
mostly depends on the halo mass, and so on its concentration (e.g., \citealt{gao04,contini12,prada12}). This result opens to the idea that the ICL cannot be simply thought as an envelope of the galaxy at which it is associated,
but rather a self-consistent component, because (by definition) its stars are not bound to any galaxy and it can be extended as far as hundreds kpc. To this, we must also add the amount of ICL formed around satellites galaxies during the so-called pre-processing (see \citealt{han18}), and the non-negligible part of it that is accreted during the growth of the group/cluster (see, e.g. \citealt{contini14}). 

During the last years, some authors attempted to study the connection between the growth of the ICL and the growth of the BCG (see references above), and only very recently by looking at the relation between 
the stellar mass of BCG and ICL within a given aperture and the halo mass (e.g. \citealt{kravtsov18,demaio18,pillepich18,demaio20}). \cite{kravtsov18}, with a limited samples of galaxy groups and clusters in the local 
universe from X-ray observations, investigated the relation between the total stellar mass within a given aperture (what they call $M_{BCG}$, but it actually includes a significant part of the ICL) and the halo mass. They find 
a slope $\alpha=0.4 \pm0.1$, not far from 0.5, i.e. the value found by \cite{gonzalez13} (who considered the total stellar mass within the virial radius) and $0.37\pm0.05$ found by \cite{demaio18}  with a sample of 23
groups and clusters at $z\sim0.4$ and considering only the stellar mass within the innermost 100 kpc. \cite{pillepich18} used the IllustrisTNG project, which is a set of cosmological magneto-hydrodynamical simulations 
of galaxy formation that were performed by using the code AREPO (\citealt{springel10}), that comprises about 4000 groups and clusters with halo mass larger than $10^{13}M_{\odot}$. They looked at the $M_* -M_{halo}$
relation at redshift $z=0$ considering the stellar mass at different apertures: 30 kpc, 100 kpc and 2 times the radius the encloses half of the stellar mass. They found an increasing value of the slope $\alpha$ with 
increasing aperture that goes from 0.49 (30 kpc), 0.59 (100 kpc) and 0.74 ($2r_{half}^{stars}$), which appears slightly high compared to the previous ones. Lately, \cite{demaio20}, with the same sample used in 
\cite{demaio18}, found $\alpha=0.48\pm0.06$ in the range $10 \,kpc<r<100 \,kpc$.

The idea behind this paper is to assume a profile for the distribution of the ICL mass and, by using the aforementioned observed data to roughly set the model, compare our predictions against different observed and 
simulated properties of the BCG+ICL system. We will assume a modified version of the NFW profile (\citealt{navarro97}) to describe the distribution of the ICL mass in a halo. The ICL is mainly formed by stellar stripping 
of galaxies and mergers between them as they orbit around the center of the potential well of the cluster. By definition of ICL, its stars are not bound to any galaxy but only to the potential well of the halo, and so is dark matter.
It is reasonable to assume that the ICL would follow the gravitational potential in the same way (or similar to) dark matter does (see also \citealt{montes19}). 

We will take advantage of our model for the formation of the ICL (latest version described in \citealt{contini19}). A semi-analytic model does not provide any spatial information regarding the distribution of stars in 
galaxies. In order to have such information, we will assume a profile for the distribution of the ICL mass such that we can obtain the amount of stellar mass in ICL at any clustercentric distance. The parameter space of 
the profile will be tested against the available observational data and the prediction of the model compared with a plethora of recent simulated/observed radial distributions of BCG/ICL mass.

The paper is organized as follows. In Section \ref{sec:methods} we briefly describe our model and the profile used to distribute the ICL mass in haloes . In Section \ref{sec:results}  we present our analysis, which 
will be fully discussed in Section \ref{sec:discussion}, and in Section \ref{sec:conclusions} we summarize our main conclusions. Throughout this paper we use a standard cosmology, namely: $\Omega_{\lambda}=0.76$, 
$\Omega_{m}=0.24$, $\Omega_{b}=0.04$, $h=0.72$ , $n_s =0.96$ and $\sigma_{8}=0.8$. Stellar masses are computed with the assumption of a \cite{chabrier03} Initial Mass Function (IMF), and all units are $h$ 
corrected. In the rest of the paper we refer to $R_{200}$ and $R_{500}$ (depending on the particular results we are comparing our predictions to) as the radii that enclose a mean density of 200/500 times the critical density of the Universe at the redshift of interest, and similarly for the masses enclosed to them, $M_{200}$ and $M_{500}$.

\section[]{Methods}  
\label{sec:methods}

We take advantage of the semi-analytic model developed in \cite{contini14}, which is a modified version of the one described in \cite{delucia07}, and that considers different mechanisms for the formation and evolution of 
the ICL, mainly stellar stripping and galaxy mergers. The full prescription has been further improved in \cite{contini18} and \cite{contini19}, so we refer the reader to these papers for the detail of the modelling, while below 
we provide a summary of the main features and assumptions.

\subsection[]{General Features of the Semi-Analytic Model}

The semi-analytic model ran on the mergers trees of a set of high-resolution N-body simulations whose details are provided in \cite{contini12} and \cite{contini14}. The set comprises 27 zoom-in cluster simulations whose 
data have been stored at 93 output times, spanning a redshifts range between $z=60$ and the present day, by using the cosmology reported in Section \ref{sec:intro}. At $z=0$ our sample counts 361 groups and clusters
with $M_{200}$ larger than $10^{13} \, M_{\odot}/h$ and up to more than $10^{15} \, M_{\odot}/h$. Given the fact the mass of each dark matter particle is $10^8 \, M_{\odot}/h$, we resolve the largest clusters with more than 
10 million particles.

To identify subhaloes within main haloes, we took advantage of the algorithm {\small SUBFIND} (\citealt{springel01}). This algorithm decomposes each FOF group into a set of different substructures that are identified 
as local overdensities in the field of the background halo. Similarly to previous works, where we used the same set of simulations, we considered as genuine subhaloes only those that could retain at least 20 bound particles.
{\small SUBFIND} has been found to have some difficulties in accurately recover subhaloes particles when they are very close to the halo centre (see, e.g., \citealt{muldrew11}), which in principle can affect the predictions of 
the amount of ICL. However, our prescription for the formation of the ICL (see below in this section for a detailed discussion) acts almost entirely on galaxies for which their subhaloes went under the resolution of the 
simulation (20 particles), so that we do not expect this limitation of {\small SUBFIND} to influence the amount of ICL formed. Nevertheless, altering the number of bound particles required for a subhalo to be considered a
genuine one and/or altering the mass resolution (higher/lower particle mass) do have an effect on the amount of ICL formed, simply because it affects the number of galaxies allowed to experience stellar stripping. For quantitative and qualitative details on this point we refer the reader to \cite{contini14}, where the issue has been fully discussed in a dedicated appendix.

The semi-analytic model populated subhaloes with galaxies by means of several physical mechanisms that describe our current knowledge of galaxy formation and evolution, and by using the information on dark matter stored in the trees. Among all, the model includes the treatment of gas cooling, star formation, SN and AGN feedback, and the formation of the ICL which is the most important prescription in the context of this paper.

To implement the formation of the ICL we use the model named "Tidal Radius" in \cite{contini14}, which operates on satellite galaxies, both with and without dark matter component. For each satellite, the semi-analytic model derives a tidal radius, $R_t$,  which depends on the mass of the galaxy itself, the mass of the main halo, and the distance between the galaxy and the center of the halo where the BCG is positioned. Every galaxy, including 
satellites, is assumed to be a two-component system with a spheroidal bulge and an exponential disk. If the tidal radius $R_t$ is small enough to be contained within the bulge, we assume the satellite galaxy to be disrupted and all
its stellar mass moved to the ICL component associated with the BCG. If the tidal radius $R_t$ is larger than the bulge radius but smaller than the disk radius $R_{sat}$, we assume that the mass in the shell $R_t -R_{sat}$
is stripped and added to the ICL component of the BCG. 

The implementation just described is directly applied to "orphan" galaxies, i.e. satellites that have lost their dark matter component (or it went under the resolution of the simulation), while for satellites with a dark matter component there is an extra requirement:  we impose that the half mass radius of the dark matter subhalo that contains the satellite has to be smaller than the half mass radius of the disk component before stellar stripping can start.
The Tidal Radius Model considers only the stripping of material while satellites are orbiting around the potential well of the cluster. The semi-analytic model allows for another channel for the formation of the ICL and it is 
provided by mergers, both minor and major.  At each merger between a satellite and a central galaxy, 20\% of the stellar mass of the satellite is moved to the ICL component associated with the central, and the rest added to 
its stellar mass (the reason of the percentage chosen is explained in \citealt{contini14}).  

Other two important assumptions must be noted and concern the amount of ICL that central galaxies, i.e. BCGs included, can acquire in a non-direct way from satellite galaxies. Once a satellite with a dark matter 
component experiences the first episode of stripping, we assume that its ICL component is accreted to the ICL component associated with the current central galaxy.  Similarly, once satellites become orphans or are 
accreted into larger haloes, their ICL component (if any) is moved to the ICL component of the current central galaxy. In both cases the ICL has been already formed and we usually refer to these processes as pre-processing,
in the sense that the ICL formed somewhere else and then accreted by the central galaxy.

\subsection[]{ICL Profile}
\label{sec:profile}

For the purpose of our study, we need to model the mass distribution of the ICL. The main reason for that is given by the fact that a semi-analytic model does not provide any spatial information but the position of the 
galaxies (or others such as bulge/disk radii). A semi-analytic model is conceptually different from a hydrodynamical simulation, which works with particles that can be followed one by one. In order to describe the spatial 
distribution of the stars that constitute the ICL, we assume a modified version of the NFW profile. As anticipated in Section \ref{sec:intro}, such a profile is a reasonable assumption considering that the stars belonging to 
the ICL are not bound to any galaxy but only to the potential well of the cluster and so, like dark matter particles, they can be considered as a collisionless system. A similar argument has been recently discussed in 
\cite{montes19} (but see also \citealt{alonso20,kluge20}). 

The NFW profile reads as follow:
\begin{equation}\label{eqn:nfw}
\rho (r) = \frac{\rho_0}{\frac{r}{R_s}\left(1+\frac{r}{R_s}\right)^2} \, ,
\end{equation}
where $\rho_0$ is the characteristic density of the halo at the time of its collapse and $R_s$ is its scale radius, i.e. the radius at which the slope of the profile ($\log (\rho)$) is equal to -2.  Once the scale radius is linked 
to the virial radius $R_{200}$, it is possible to define the concentration of the halo as 
\begin{equation}\label{eqn:concentration}
c = \frac{R_{200}}{R_s} \, .
\end{equation}
We modify the classic profile described in Equation \ref{eqn:nfw} by introducing a new parameter in Equation \ref{eqn:concentration}. Basically, we define the ICL concentration as
\begin{equation}\label{eqn:gamma}
c_{ICL} = \gamma \frac{R_{200}}{R_s} \, .
\end{equation}

The new parameter $\gamma$ could be, in principle, a function of several halo properties and/or redshift. For the sake of simplicity, we will consider this parameter to be halo and redshift independent, but we will show 
our results by testing several values. Under this assumption, $\gamma$ is then a multiplicative factor which acts on the concentration of the dark matter halo and returns the concentration of the ICL mass distribution. 
Considering the baryonic nature of the ICL, the channels of its formation and evolution, it is likely that the ICL is more concentrated than the dark matter halo, which translates in $\gamma > 1$. This is 
supported by several recents theoretical works (e.g. \citealt{contini18,pillepich18}) and observational studies (e.g. \citealt{montes19}). 

In the next section we will test the ICL profile described by Equations \ref{eqn:nfw} and \ref{eqn:gamma} . The new profile allows us to know the ICL mass within any radial bin, which means that we can overplot our 
predictions on observational (and theoretical) results to test the validity of the profile itself, by varying the value of $\gamma$. We will focus in particular on the BCG+ICL mass within 100 kpc at different redshifts, and 
after having set the values of $\gamma$ (we remind the reader that $\gamma$ can be a function of redshift or some halo properties), we will test our predictions on several other observed and simulated quantities.

\section{Results}
\label{sec:results}

\begin{figure*} 
\begin{center}
\begin{tabular}{ccc}
\includegraphics[scale=.33]{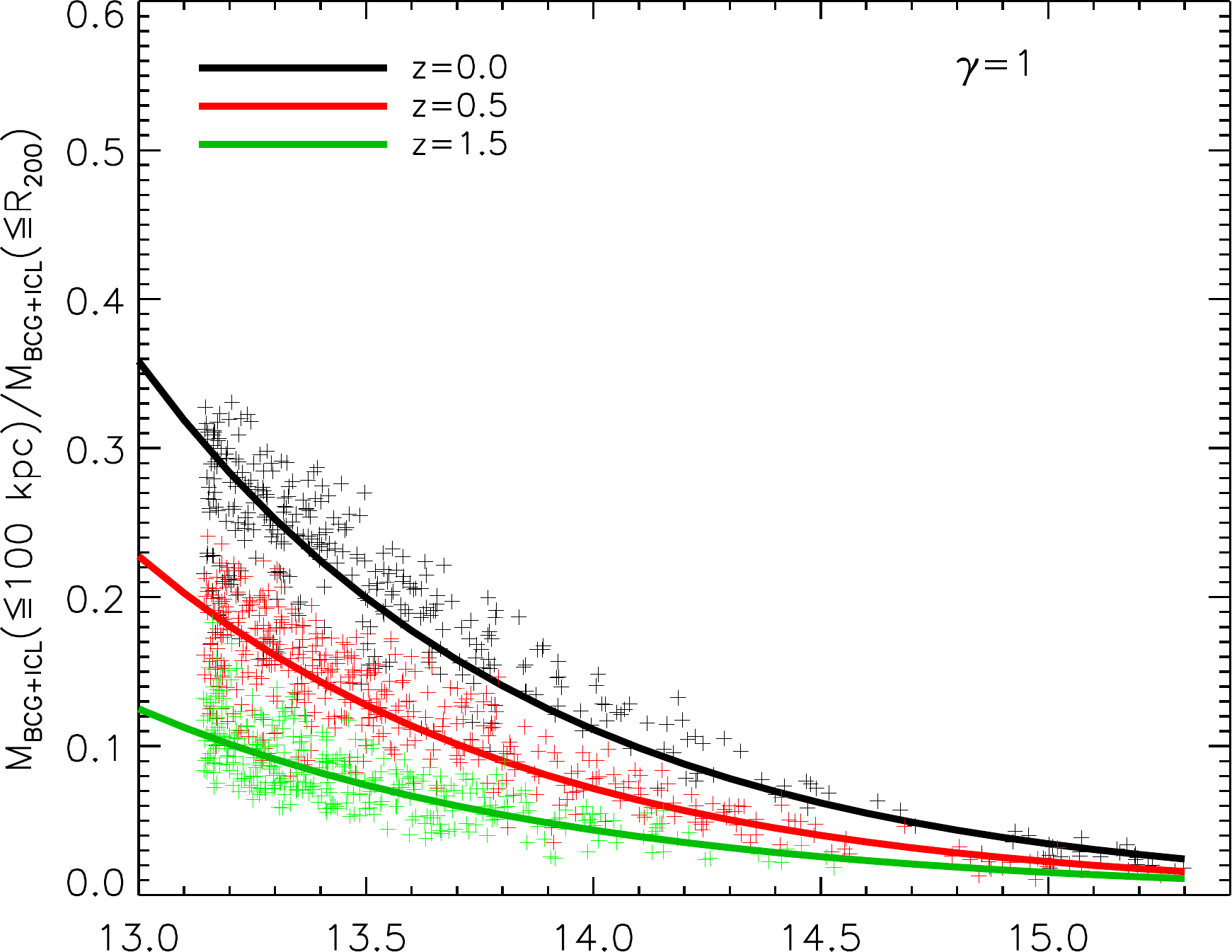} &
\includegraphics[scale=.33]{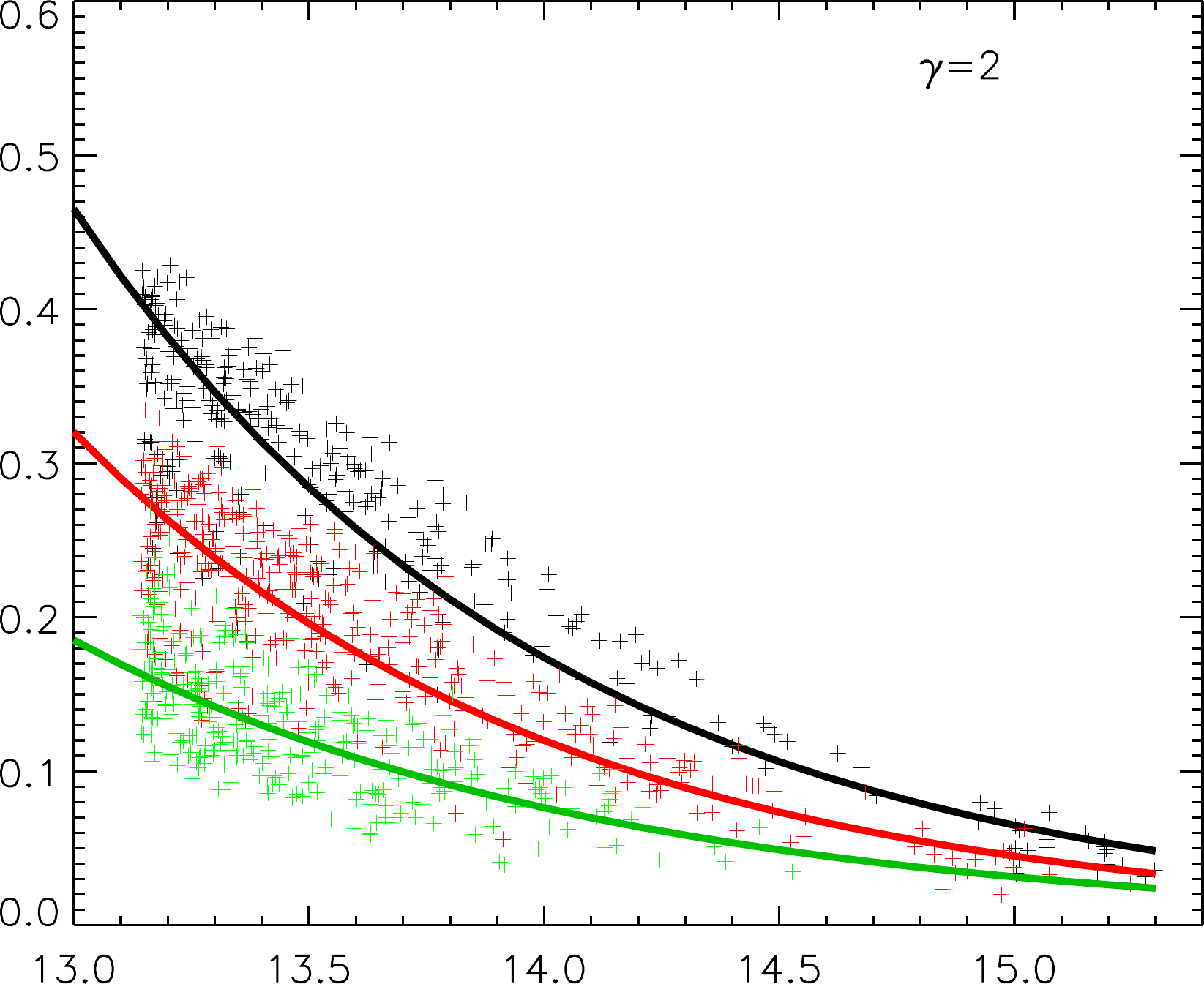} &
\includegraphics[scale=.33]{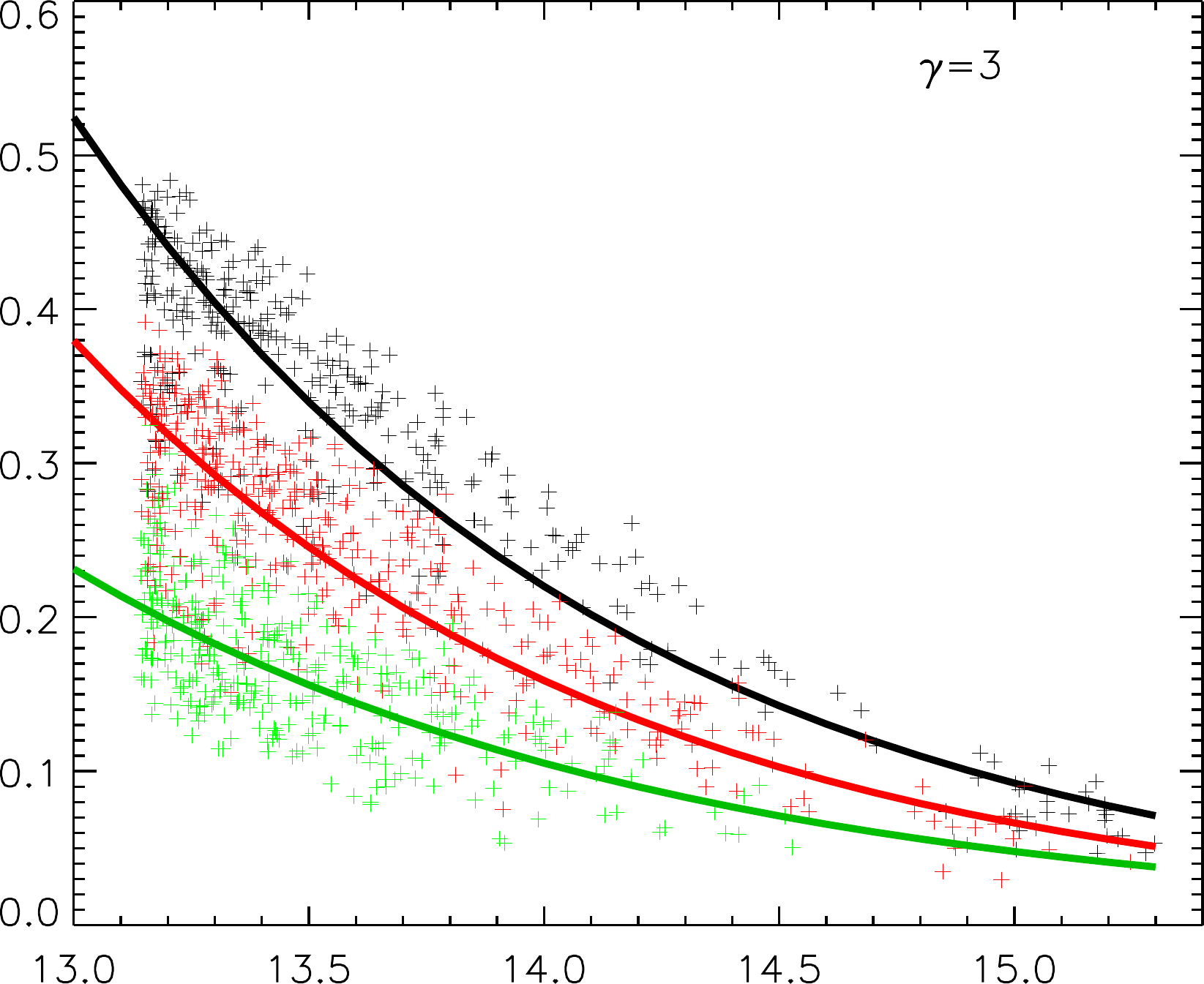} \\
\includegraphics[scale=.33]{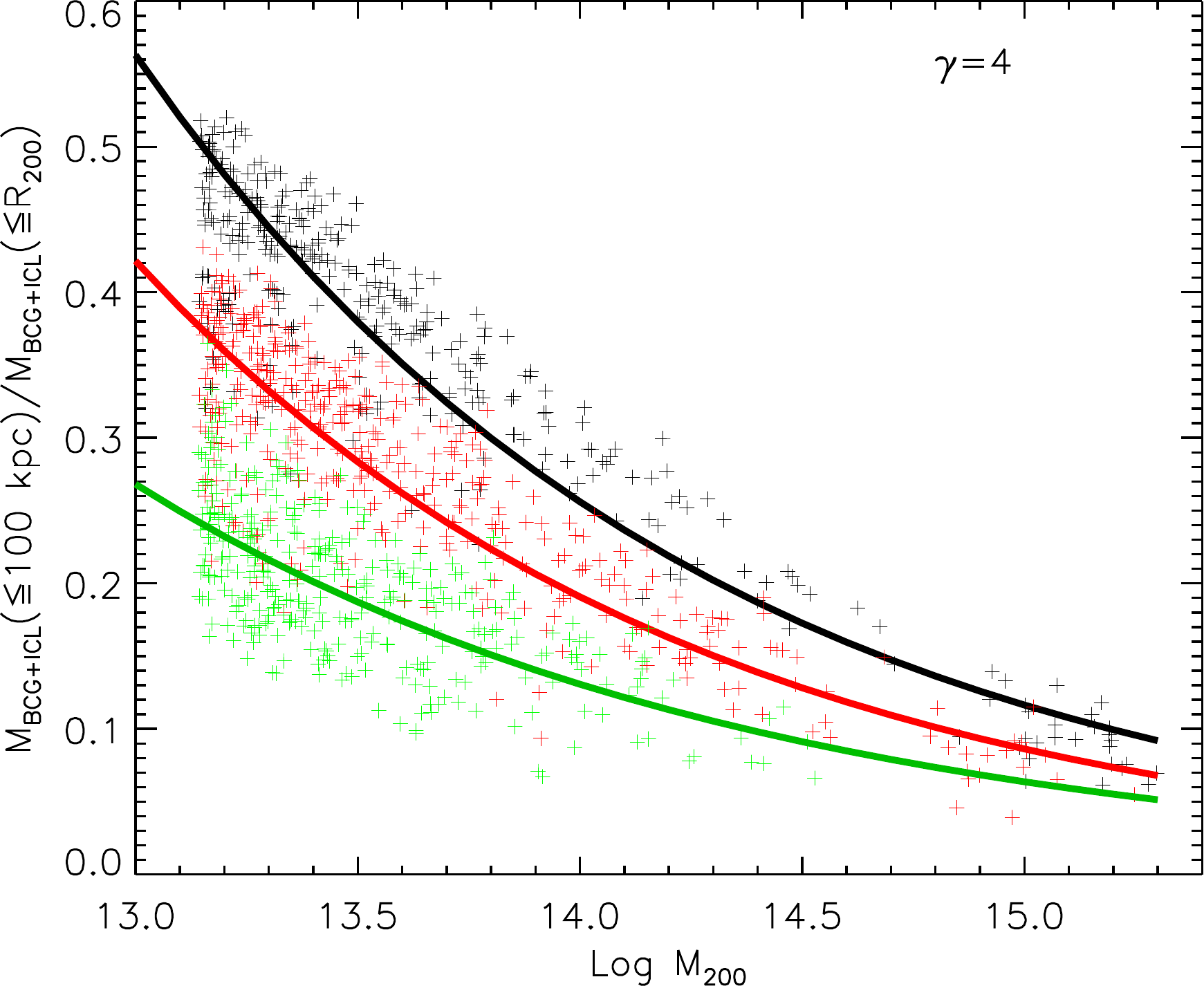} &
\includegraphics[scale=.33]{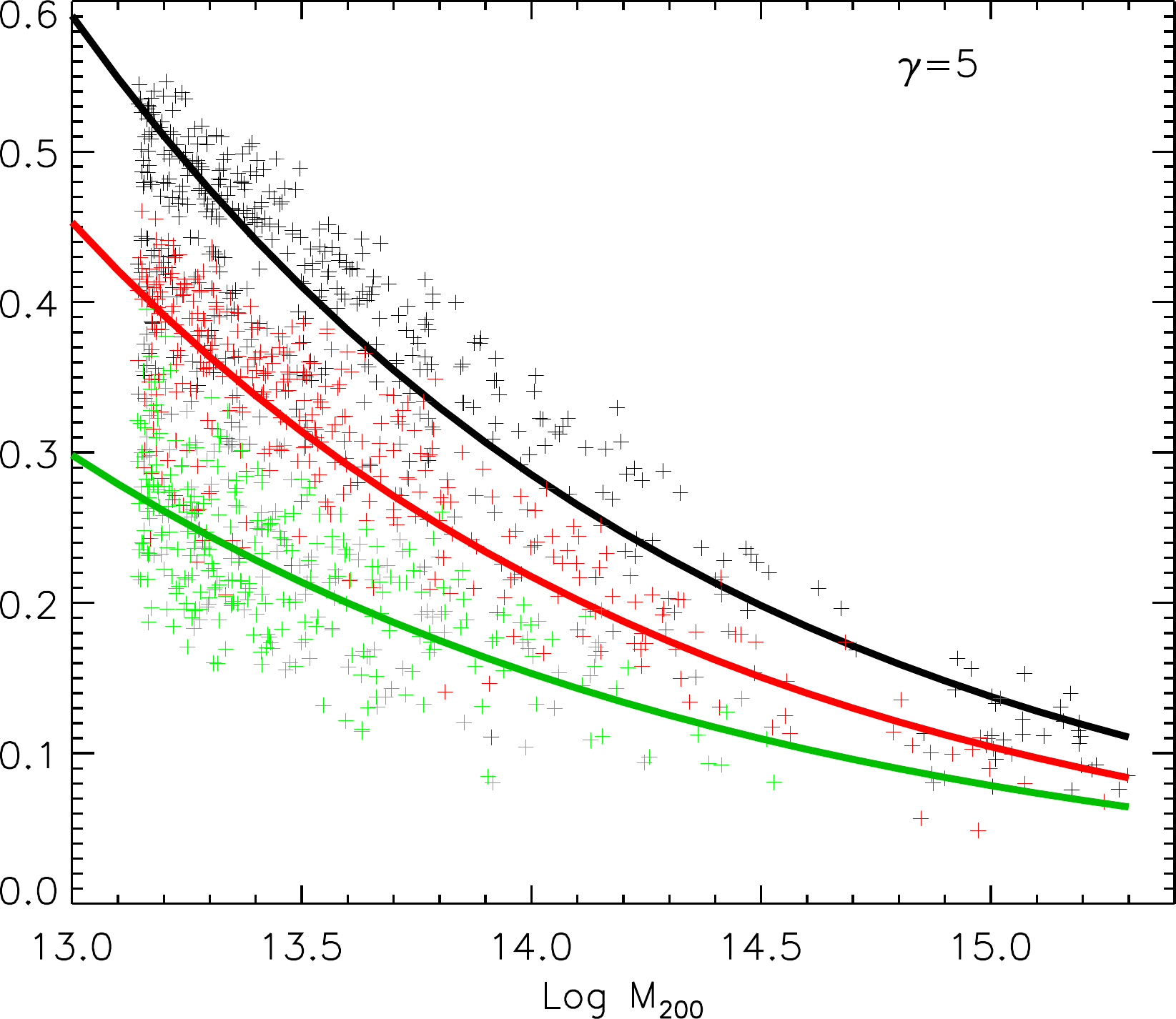} &
\includegraphics[scale=.33]{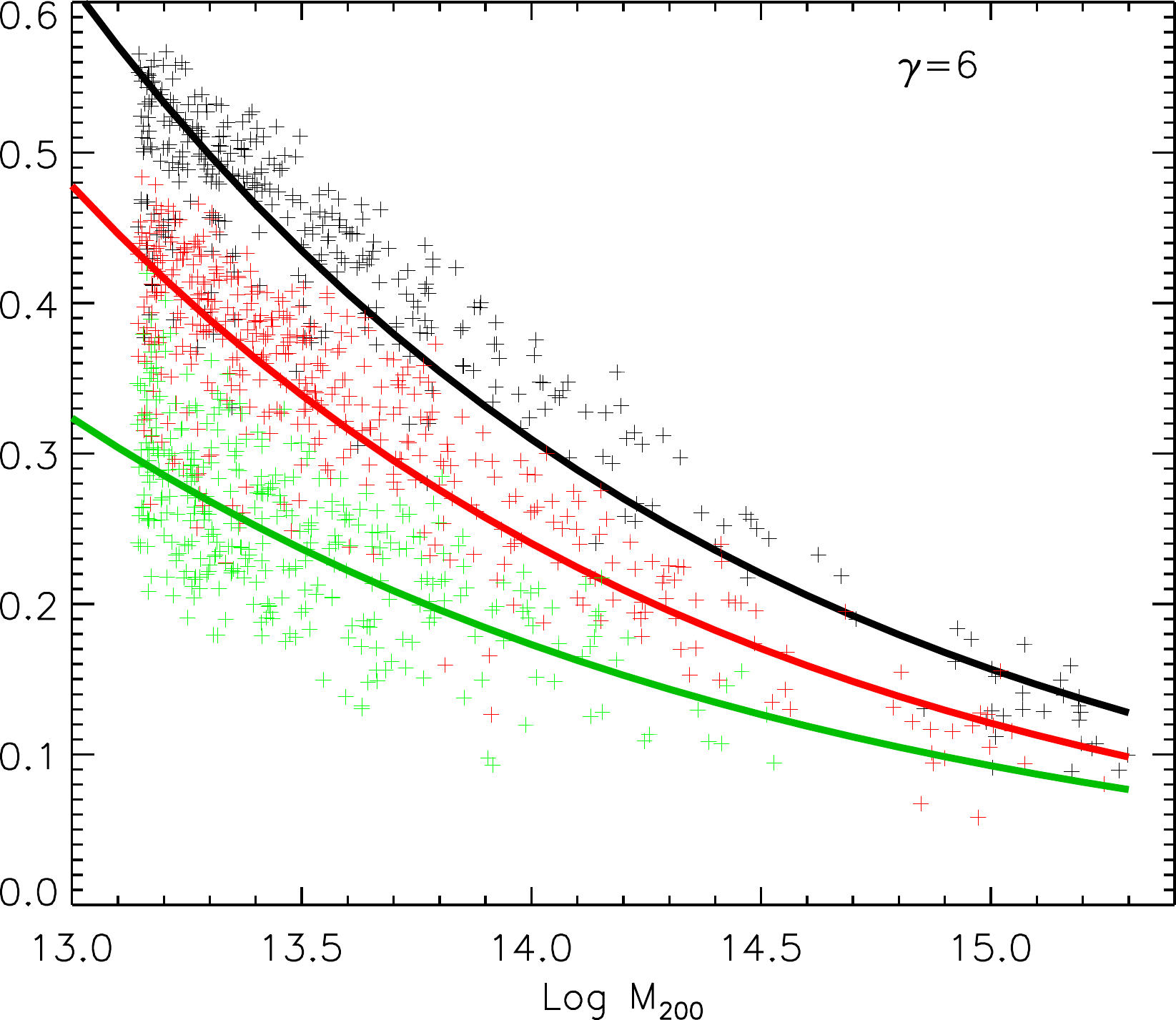} \\
\end{tabular}
\caption{Ratios between the ICL-BCG mass within 100 kpc from the halo center and the total ICL-BCG mass within $R_{200}$, as a function of halo mass and 
at different redshifts as reported in the legend.  Each panel shows the prediction of the model for a particular value of $\gamma$, which ranges from one to six. The percentage 
of the ICL-BCG mass enclosed in 100 kpc is an increasing function of redshift, (different colors), an increasing function of $\gamma$ (different panels), and a decreasing function 
of halo mass. These trends are in line with the fact that less massive haloes are more concentrated and smaller than more massive ones, and with the fact that haloes of a given 
mass are, on average, less concentrated at higher redshift.}
\label{fig:massratio}
\end{center}
\end{figure*}

For the purposes of our analysis, we  apply the model described in Section \ref{sec:profile} at three redshifts, namely: z=0, 0.5, 1.5.  Before going to the detail of the analysis, in Figure  \ref{fig:massratio} 
we show the ratio between the mass in ICL and BCG contained within 100 kpc \footnote{In this paper we do not go below 100 kpc, and consider the mass of the BCG to be fully contained 
within that distance from the halo center.} from the center of the halo and the total ICL-BCG mass within $R_{200}$, as a function of the halo mass (which spans a wide range from 
$\log M_{200} \sim 13$ to $\log M_{200} \sim 15.3$), and at different redshifts (different colors) as shown in the legend. Each panel shows the prediction of the model for a particular value of the 
parameter $\gamma$, from $\gamma=1$ (top left panel), to $\gamma=6$ (bottom right panel). By focusing on any of the panels (i.e., independently of the value of $\gamma$), we can see that the 
the percentage of the mass in ICL and BCG in 100 kpc depends on both redshift and halo mass. The trend with redshift appears to be moderately strong, and so the trend with halo mass especially for 
higher values of $\gamma$. 

Let's now consider an intermediate value of $\gamma$, such as $\gamma=3$ (top right panel). In this particular case and at the present time, the percentage changes from 
50\% in the low halo mass end, to around 10\% in the high halo mass end. If we consider the highest redshift, the change of the percentage is less dramatic, from $\sim 20\%$ to $< 10\%$, 
which means that the redshift increase is also halo mass dependent. Instead, by focusing on the different panels, it is possible to see another trend: the higher the value of $\gamma$, the higher the 
percentage of ICL and BCG mass within 100 kpc, regardless the halo mass, although the effect is certainly more evident toward lower halo masses. This trend can be easily explained by the fact 
that less massive haloes are more concentrated (and smaller) than more massive ones (\citealt{gao04,prada12}), together with the fact that haloes of a given mass are also, on average, less 
concentrated at higher redshift (e.g \citealt{gao11,contini12}). 

The key-point of this figure is that $\gamma$ plays an important role and in the rest of the following analysis we set the model to the value that best fits the observations, according to the redshift of the 
observed data. We remind the reader that the parameter $\gamma$ can be a function of halo mass, but can in principle be also a function of redshift, i.e. at different redshifts, different values of $\gamma$ 
can be more appropriate.

\begin{figure*} 
\begin{center}
\begin{tabular}{cc}
\includegraphics[scale=.98]{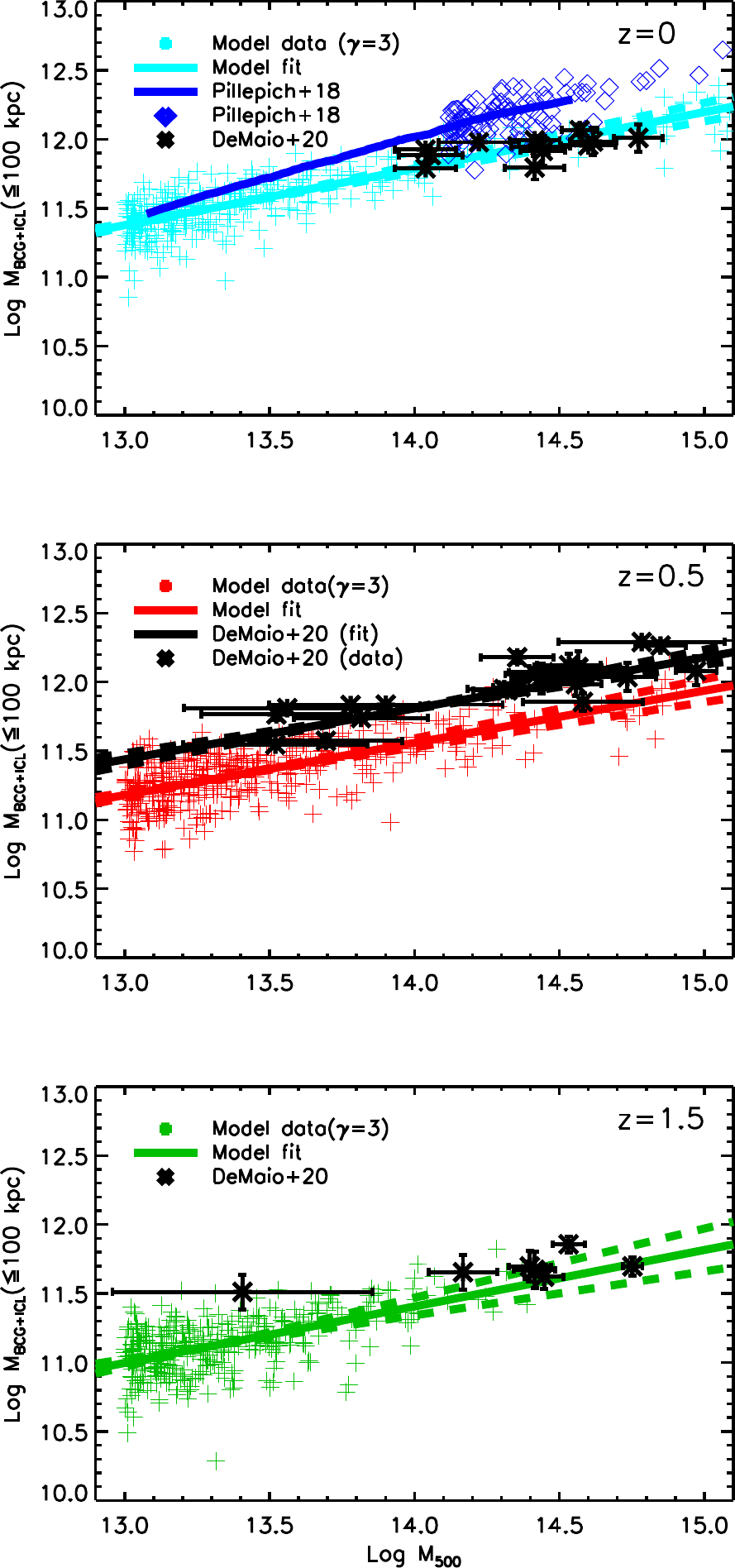} &
\includegraphics[scale=.98]{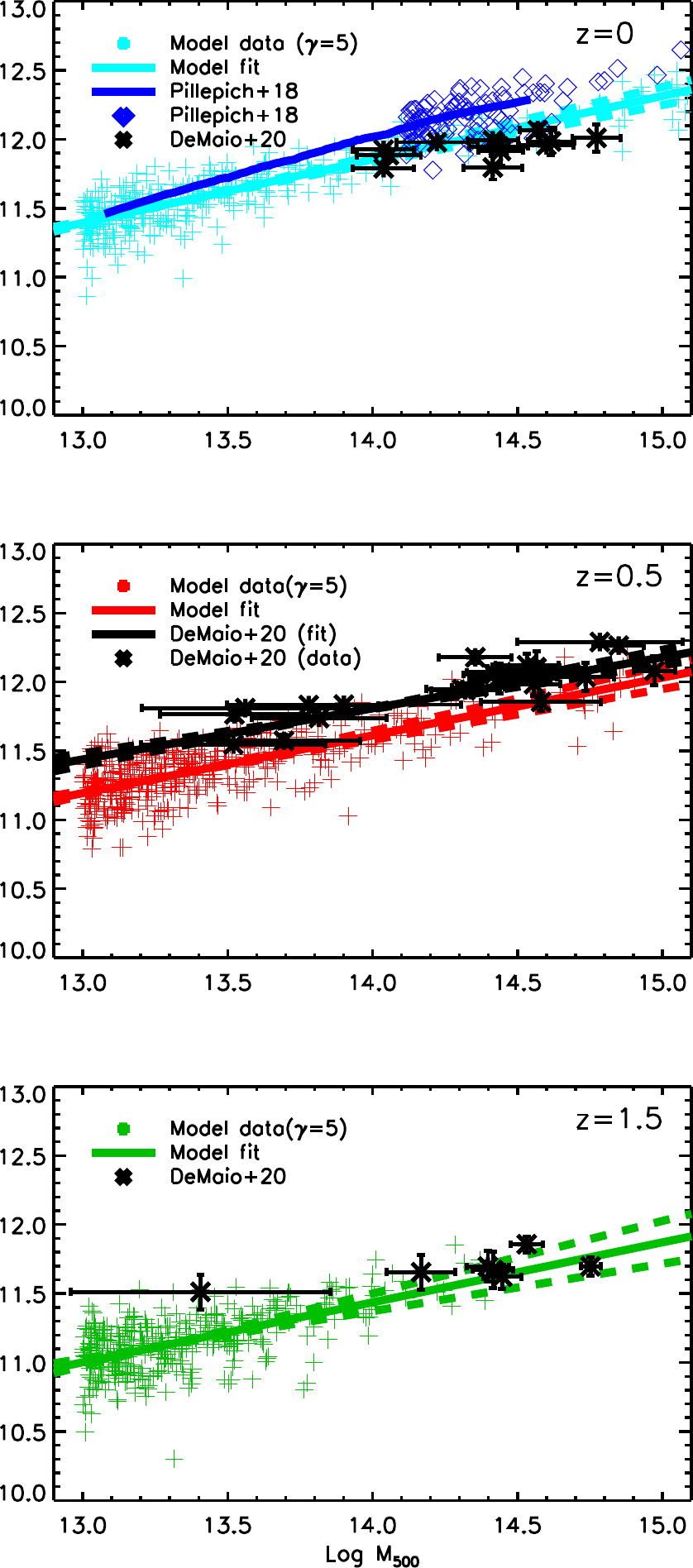} \\
\end{tabular}
\caption{Relation between the ICL-BCG stellar mass enclosed in 100 kpc and the halo mass $M_{500}$, for groups and clusters at different redshifts as indicated in the legend.
The two columns refer to the model predictions for $\gamma =3$ (left column) and $\gamma =5$ (right column). The model predictions are compared with the observed set of data by 
\cite{demaio20} and the results of the IllustrisTNG simulations (\citealt{pillepich18}).  Our model agrees well with the observed data in the local Universe and better than the prediction
of the simulations by preferring low values of $\gamma$. At higher redshift it reproduces fairly well the observed trend and prefers higher values of $\gamma$. The values of the intercepts 
and slope of each fit are reported in Table \ref{tab:tab1}.}
\label{fig:m100}
\end{center}
\end{figure*}

\begin{table}
\caption{Intercepts and slopes with scatters of the $M^*_{100}-M_{500}$ relation (Figure \ref{fig:m100}) at the redshifts investigated, for $\gamma=3$ and $\gamma=5$. Data points have 
been fitted with the simple linear fit $\log M^*_{100}=\alpha \log(M_{500}/2\cdot10^{14})+\beta$.}
\begin{center}
\begin{tabular}{llllll}
\hline
z & $\beta$ ($\gamma=3$) & $\alpha$ ($\gamma=3$) & $\beta$ ($\gamma=5$) &  $\alpha$ ($\gamma=5$)\\
\hline
0       &  $11.91\pm0.04$ & $0.41\pm0.04$ & $12.00\pm0.04$ & $0.46\pm0.04$ \\
0.5    &  $11.67\pm0.05$ & $0.38\pm0.05$ & $11.74\pm0.05$ & $0.42\pm0.05$ \\
1.5    &  $11.53\pm0.09$ & $0.41\pm0.09$ & $11.57\pm0.09$ & $0.44\pm0.09$ \\
\hline
\end{tabular}
\end{center}
\label{tab:tab1}
\end{table}

Figure \ref{fig:m100} shows the relation between the stellar mass of ICL and BCG within 100 kpc, and the mass $M_{500}$ (hereafter $M^*_{100}-M_{500}$), for haloes at different redshift,
from $z=0$ (top panels) to $z=1.5$ (bottom panels). Our model predictions are compared with those of the IllustrisTNG simuations (\citealt{pillepich18}) at $z=0$, and with the observed data 
by \cite{demaio20} at all redshifts investigated (see legend). The two columns refer to the model predictions where we used $\gamma=3$ (left column) and $\gamma=5$ (right column).  We have 
fit our data points with the simple linear fit $$\log M^*_{100}=\alpha \log(M_{500}/2\cdot10^{14})+\beta$$ and values of intercepts, slopes and scatter are reported in Table \ref{tab:tab1}. Let's 
comment each panel of the figure. The first line of panels refers to the present time, and as for the others, the only difference between the two is given by the different values of $\gamma$ (3 or 5) used. 
At redshift $z=0$ our model (cyan lines and symbols) agrees well with the observational data (black symbols), and even better than the prediction of the IllustrisTNG 
simulation (blue line and symbols), by preferring a low value of $\gamma$. At redshift $z=0.5$, although the model (red lines and symbols) reproduces well the slope of the the observational data
(black lines and symbols), it is biased low with respect to them.  What is possible to note at this redshift is that observations suggest higher values of $\gamma$ with respect to $z=0$. The trend is confirmed at much higher 
redshift (bottom panels), where a simple linear fit (green lines) of our model predictions (green symbols) is able to match most of the observed data (black symbol). A possible reason for the bias seen
at higher redshift can be the wide redshift range investigated in the work of \cite{demaio20}, where $z\sim 0.4/1.55$ are just the median of the redshift distributions. Another possible reason can be
that the observed BCG or ICL masses are intrinsically different from those derived by the model  \footnote{A higher $\gamma$ can transfer more ICL mass within 100 kpc. We ruled out 
this possibility by testing several extreme values.}. We will come back on this point in Section \ref{sec:discussion}, while for the rest of our analysis we focus on the present time with the choice of $\gamma=3$.

\begin{figure}[hbt!]
\includegraphics[scale=0.47]{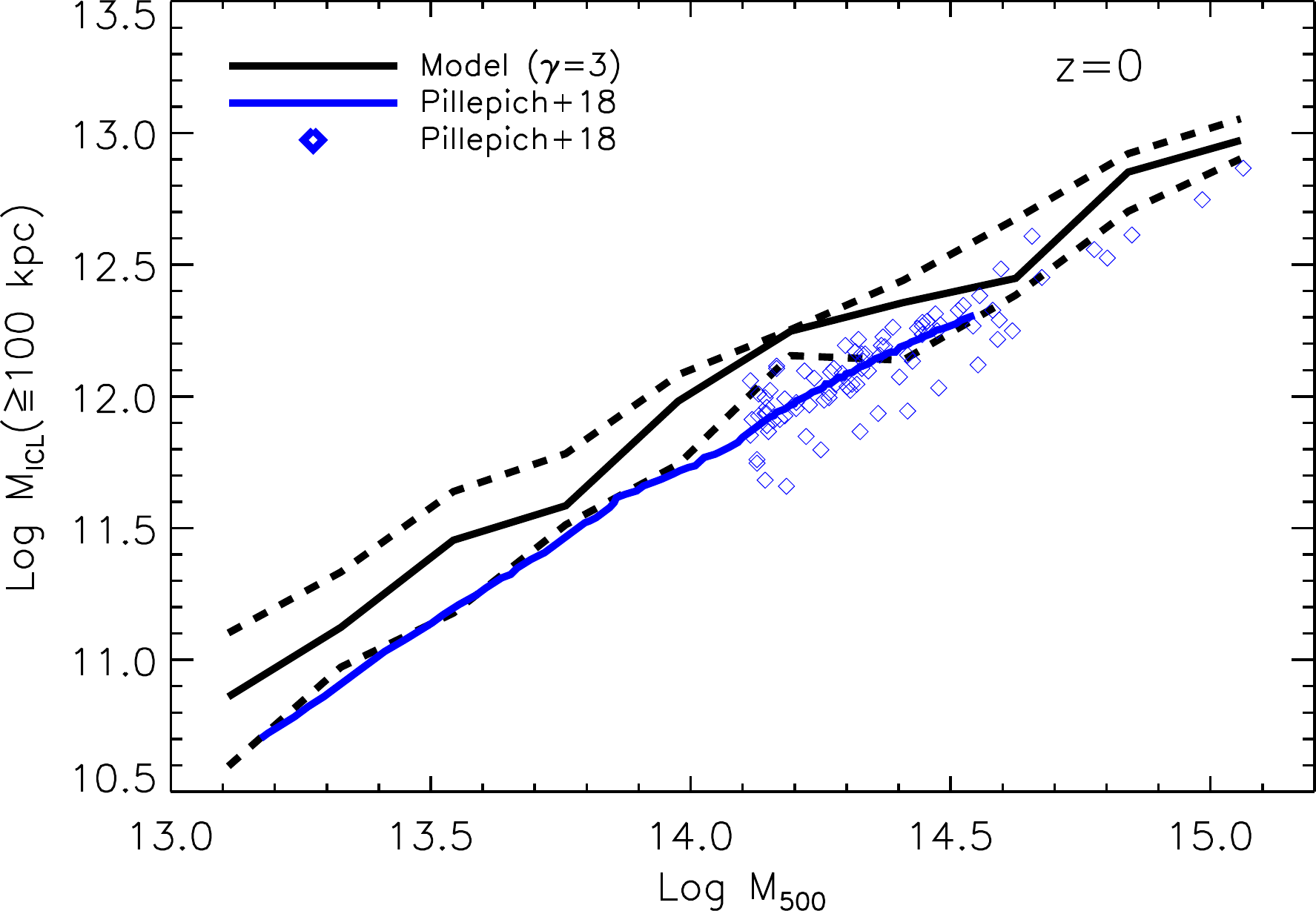}
\centering
\caption{Mass of the ICL outside 100 kpc as a function of the halo mass $M_{500}$ at $z=0$ and $\gamma =3$, compared with the results of \cite{pillepich18}. Although the trend found by 
\cite{pillepich18} is reproduced, our model (black lines for median, 16th and 84th percentiles of the distribution) predicts slightly more ICL outside 100 kpc with respect to their simulation.}
\label{fig:icl100_halom}
\end{figure}

In Figure \ref{fig:icl100_halom} we focus on the mass of ICL outside 100 kpc, and we plot it as a function of the mass $M_{500}$, at $z=0$. Our model predictions (black lines) are compared with the 
result of \cite{pillepich18} (blue line and symbols). The trend found in the analysis of the IllustrisTNG data is reproduced, but our model predicts slightly higher amounts of ICL outside 100 kpc than those 
predicted by the simulation. Roughly speaking, IllustrisTNG data are $1-\sigma$ lower than our predictions. A caveat must be noted. \cite{pillepich18} call "central" all the stellar mass within a given distance
from the center, and ICL the stellar mass outside that given distance (in this case 100 kpc). Our definitions of ICL are not fully comparable because we have made the assumption that the BCG is confined 
within 100 kpc, while in principle, in \cite{pillepich18}, where no attempt to separate the two components is made, part of the BCG mass can extend farther than 100 kpc. The two definitions become the same 
in the case that no mass of their BCGs extends farther than 100 kpc. Further details on this will be given in Section \ref{sec:discussion}. Moreover, another point must be clear: both here and in \cite{pillepich18}
the assumption of spherical symmetry has been implicitly made and, as \cite{pillepich18} rightly state in their paper, it is a clear simplification.

\begin{figure}[hbt!]
\begin{tabular}{cc}
\includegraphics[scale=.47]{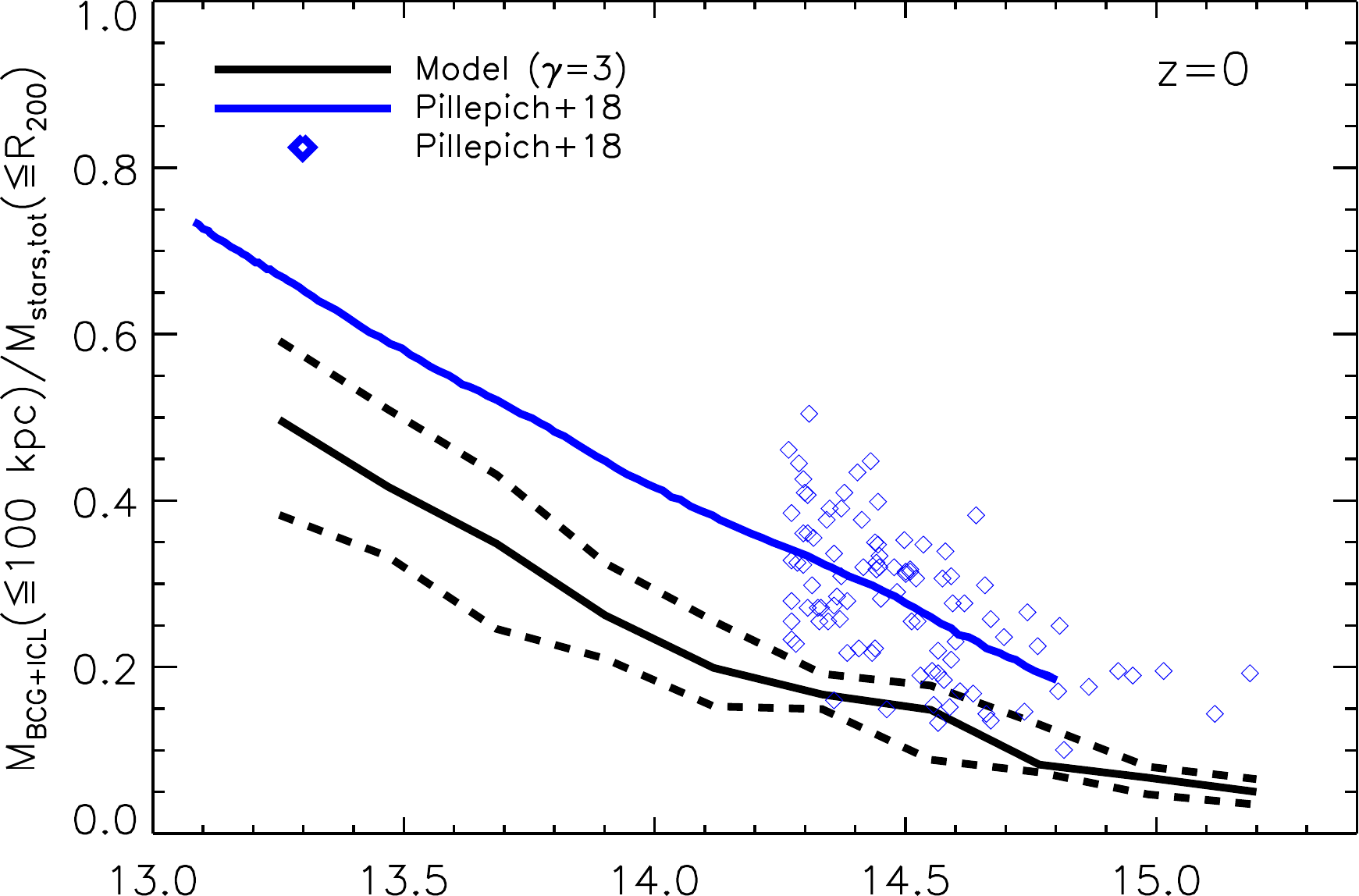} \\
\includegraphics[scale=.47]{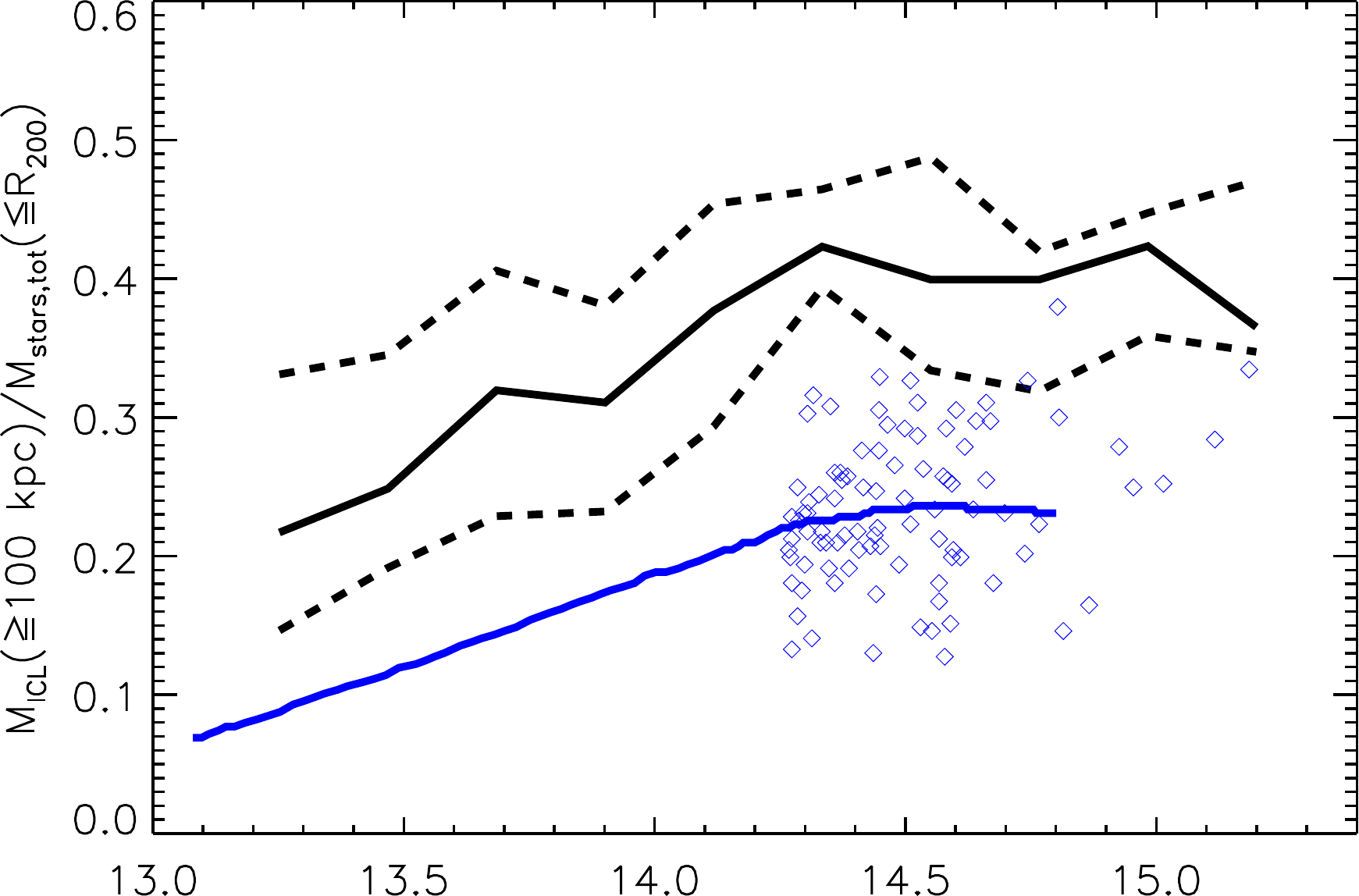} \\
\includegraphics[scale=.47]{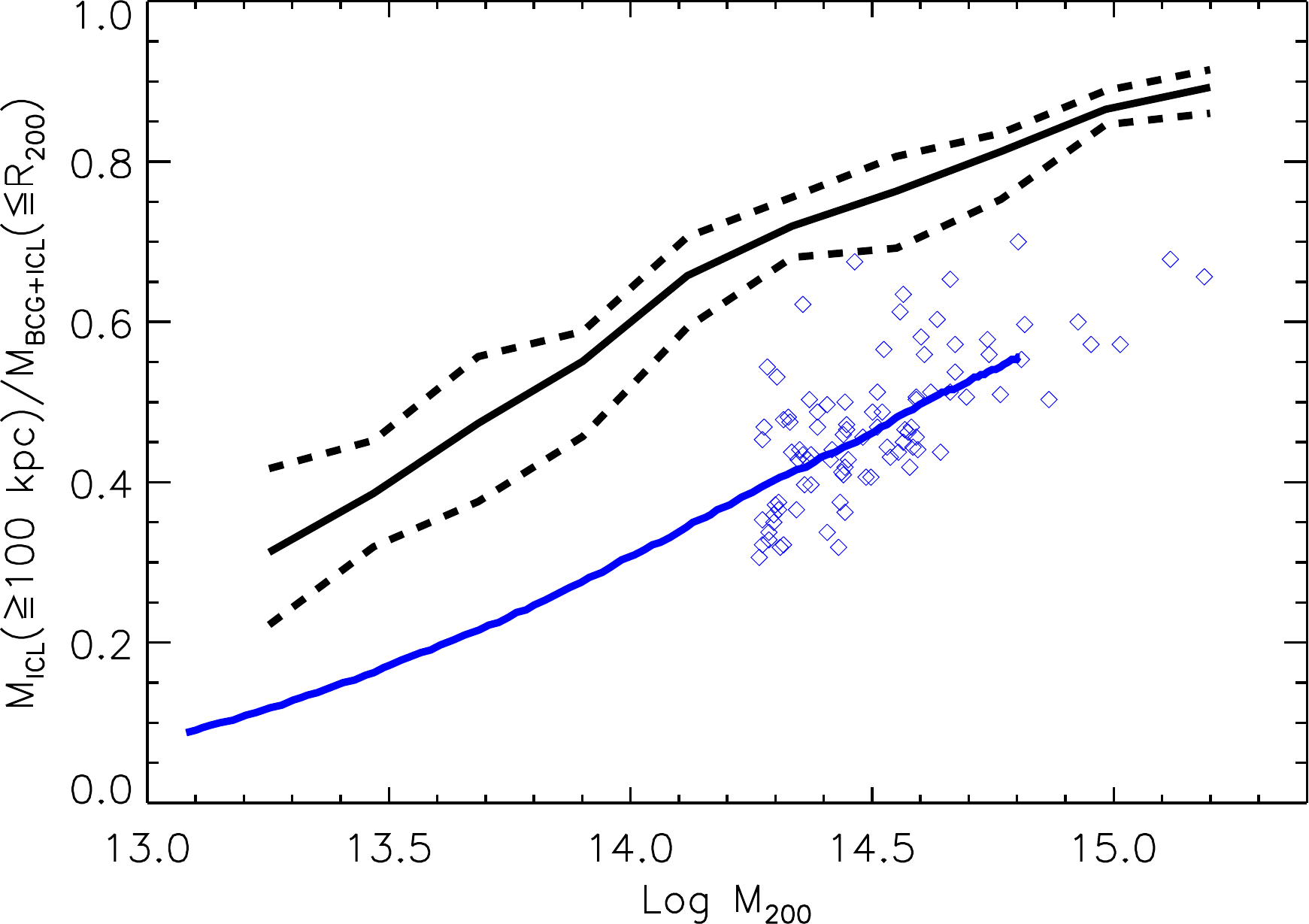} \\
\end{tabular}
\caption{Different mass ratios as a function of the halo mass $M_{200}$. From top to bottom: the mass of BCG+ICL enclosed within 100 kpc (top panel), the mass in ICL outside 100 kpc (central panel) over the total stellar 
mass within $R_{200}$, and the mass in ICL outside 100 kpc over the BCG+ICL mass within $R_{200}$ (bottom panel), as a function of the mass $M_{200}$. Our model predictions (black lines as in Figure 
\ref{fig:icl100_halom}) are compared with the predictions of the IllustrisTNG simulation (\citealt{pillepich18}). Compared to \cite{pillepich18}, our model predicts a smaller fraction of BCG+ICL mass in the innermost 100 kpc, and a higher amount of ICL in the rest of a cluster (as seen in Figure \ref{fig:icl100_halom}).}
\label{fig:fraction}
\end{figure}

In order to investigate on the possible causes of this difference, in Figure \ref{fig:fraction} we make a one-to-one comparison between our predictions and the result of the IllustrisTNG by looking at (from the
top to the bottom): the mass of BCG+ICL within 100 kpc (top panel), the mass in ICL outside 100 kpc (central panel) over the total stellar mass within $R_{200}$, and the mass in ICL outside 
100 kpc over the BCG+ICL mass within $R_{200}$ (bottom panel), as a function of the mass $M_{200}$.  The fraction of ICL and BCG mass within 100 kpc predicted by our model is lower with respect to 
the simulated data, at all halo masses. In the same halo mass range (from the low mass end to $\log M_{200} \sim 14.8$ of our sample), the fractions are 50\% and 10\% from our model, 65\% and 20\% from 
the simulation.

\begin{figure*} 
\begin{center}
\begin{tabular}{cc}
\includegraphics[scale=.49]{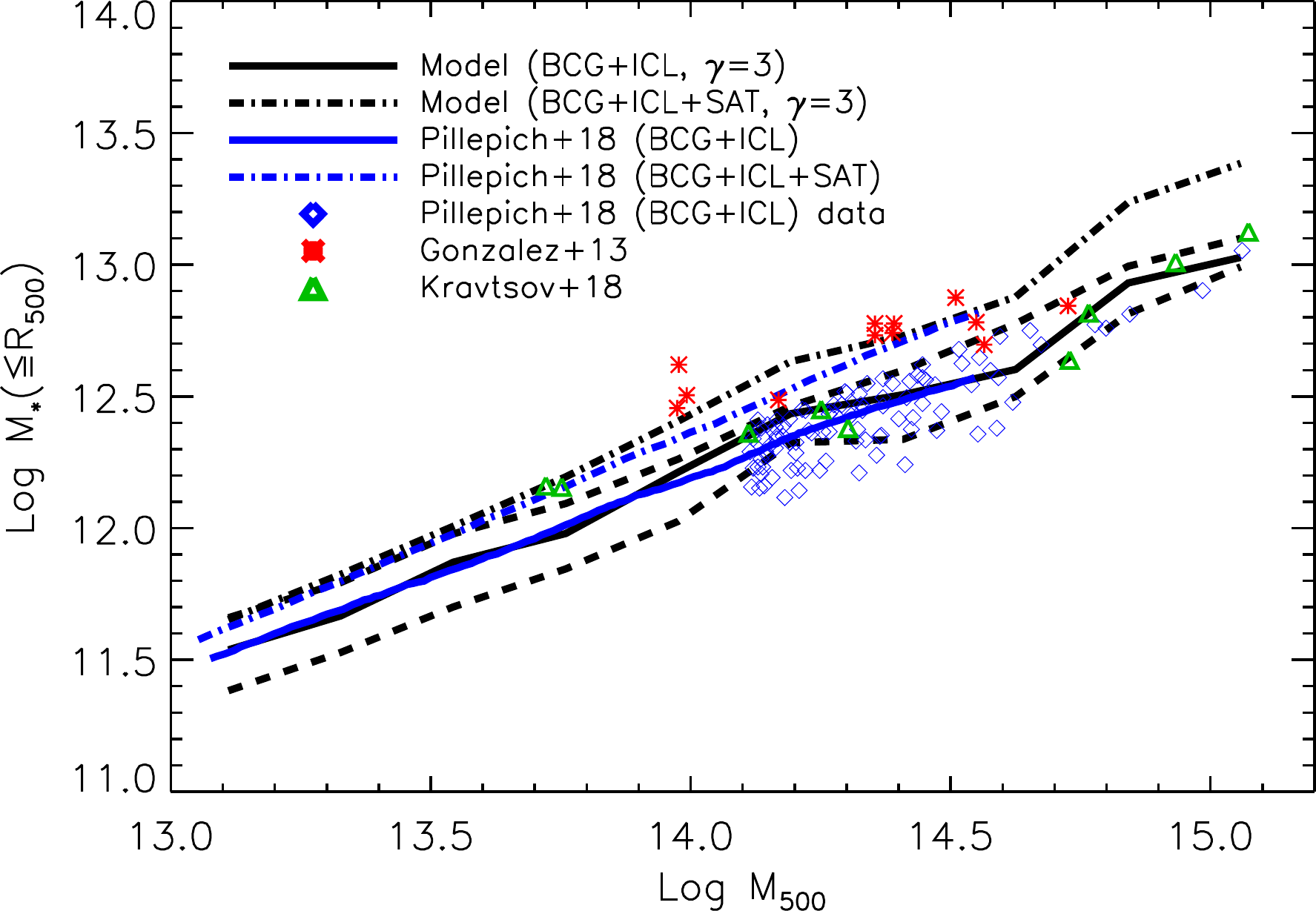} &
\includegraphics[scale=.49]{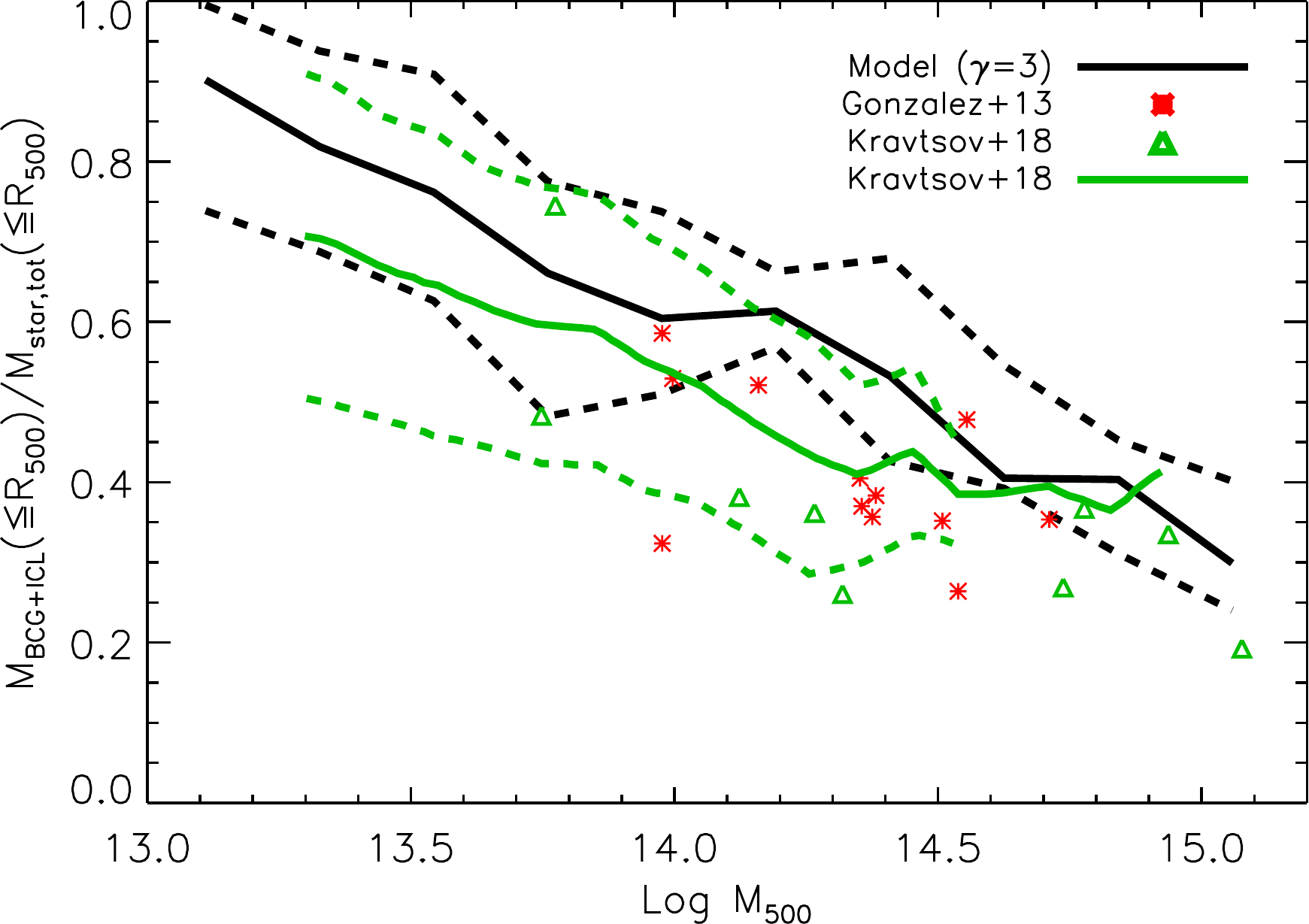} \\
\end{tabular}
\caption{Left panel: mass of BCG+ICL (black solid line) and the same with the contribution of satellite galaxies (black dash-dotted line) within $R_{500}$ as a function of the halo mass $M_{500}$ 
compared with the simulations of \cite{pillepich18} and the observational data of \cite{gonzalez13} and \cite{kravtsov18}. The amount of BCG+ICL within $R_{500}$ is in perfect agreement with the simulated results 
(and so is the total stellar mass) and in good agreement with the observed ones (even though our prediction and the simulated data are systematically lower than the observed data by \cite{gonzalez13}.
Right panel: the ratio within $R_{500}$ between the BCG+ICL and total stellar mass as a function of the halo mass $M_{500}$, at redshift $z=0$. The model prediction 
(black lines as in Figure \ref{fig:icl100_halom}) are compared with the sets of data by \cite{gonzalez13} and \cite{kravtsov18} as indicated in the legend. Overall, the observed trend is reproduced although our results appear to 
be biased with respect to the cloud of observed data. As highlighted in the text, there are caveats worth noting and they are fully discussed in Section \ref{sec:discussion}. }
\label{fig:iclbcg_halom}
\end{center}
\end{figure*}

The central panel shows the complementary plot shown by the top panel, i.e. the mass of ICL outside 100 kpc normalized to the total stellar mass within $R_{200}$. This panel clearly shows that the 
trend is inverted with respect to the results found in the top panel, i.e. our model predicts higher fractions of ICL outside 100 kpc than IllustrisTNG. As noted above, the mismatch can be due to the different 
definitions of ICL used. However, assuming that the two definitions are comparable, there are two ways to explain the non-negligible difference in the results: (a) different amount of ICL predicted by our 
model if compared with IllustrisTNG or, (b) the ICL distribute differently in IllustrisTNG. We investigate further on this below. In the bottom panel of Figure \ref{fig:fraction} we plot the fraction of ICL outside 
100 kpc over the total amount of BCG+ICL within $R_{200}$, as predicted by our model and by IllustrisTNG. Again, the two predictions are far from each other. In the halo mass range where they can 
be compared, our model goes from around 0.3 in low halo masses to 0.9 in the highest halo masses, while the fractions for IllustrisTNG are lower, 0.1 and 0.7 respectively.

In the left panel of Figure \ref{fig:iclbcg_halom} we show the mass of BCG+ICL (black solid line), and the same with the contribution of satellite galaxies (black dash-dotted line), within $R_{500}$ 
as a function of the halo mass $M_{500}$, and a detailed comparison with observed data from  \cite{gonzalez13} and \cite{kravtsov18}, and with IllustrisTNG (\citealt{pillepich18}) results. Interestingly, our predictions 
(whether we include satellites or not) are in perfect agreement with the results of IllustrisTNG, and also in good agreement with the observed data. Considering the precision with which our model and IllustrisTNG 
agree in this case, and considering what discussed above, this means that, although we both predict the same amount of BCG+ICL, the mass distributions of the ICL along the clusters are different. Our model,
with respect to IllustrisTNG,  predicts more ICL outside 100 kpc. 

In order to double check the validity of our model, we plot in the right panel of Figure \ref{fig:iclbcg_halom} the ratio within $R_{500}$ between the BCG+ICL and total stellar mass as a function of the 
halo mass $M_{500}$, at redshift $z=0$, and compare the model prediction (black lines) with the sets of data by \cite{gonzalez13} and \cite{kravtsov18}, as indicated in the legend. Overall, the observed trend is 
reproduced although our results appear to be biased with respect to the cloud of the observed data. In particular, our predictions are closer to the  green line, which refers to the results of \cite{kravtsov18} where, from 
a sample of haloes, stellar masses were assigned using stellar mass–halo mass relation derived using an abundance matching technique and assuming a scatter of 0.2 dex.  However, for the quantities shown in 
both panels there are caveats which are worth discussing, and we will fully address them in Section \ref{sec:discussion}.

In the next section we will discuss in detail the results presented in this section and, more important their consequences, in particular in light of the comparison with the observed data.

\section{Discussion}
\label{sec:discussion}

Our state-of-art model for the formation of the ICL is able to reproduce the total BCG+ICL mass within 100 kpc (this paper), as well as other important properties, as shown in former papers with detailed comparisons with observations. Given the semi-analytic nature of our model, it does not provide the spatial information of this component, i.e. how the ICL distributes along the cluster after it forms. We assumed a simple mass profile that links the distribution of the ICL to that of the dark matter, motivated by recent works such as \cite{montes19}, \cite{kluge20} and \cite{alonso20}. The ICL mass profile is generalized from a NFW density profile of the dark matter, and the connection between the two profiles is given by the parameter $\gamma$ introduced in Section \ref{sec:profile} that links the two concentrations. This modelling of the ICL profile allows us to know, with the 
necessary information such as the ICL mass and virial radius of the cluster, the amount of ICL at any clustercentric distance (under the assumption of spherical symmetry).

We tested our model against the observed data by \cite{demaio20} at different redshift and the simulated data at $z=0$ from the IllustrisTNG simulation (\citealt{pillepich18}) by looking at the $M^*_{100}-M_{500}$ 
relation, where $M^*_{100}$ is the total BCG+ICL mass within 100 kpc from the cluster center. Our profile is able to capture the slope of the relation from $z=1.5$ to the present time, while the intercept at 
$z=0.5$ seems to be lower than observed by around 0.2 dex. Overall, mild variations of  $\gamma$ do not considerably change the results, although our analysis suggests that $\gamma$ might be an increasing 
function of redshift.  In Section \ref{sec:results} we briefly discussed the gap at $z=0.5$ between the observed and predicted intercepts as possibly due to the fact that while our relation is plotted exactly at that 
redshift, observed data span a wide redshift range around the median $z\sim0.4$. It is, however, more likely that the BCG+ICL observed masses are intrinsically higher than those predicted by our model, which is
in part caused by the intrinsic difficulty in detecting the full light and/or problems in separating the source from the sky, especially at higher redshift, and in part it might be due to the fact that the model slightly 
underestimates the total BCG+ICL mass with the same degree in the whole halo mass range investigated. We will come back on this below.

The key point that comes from the comparison with both observed data and simulations at the present time (where data are supposed to be safer) is that our model is in better agreement with the observed data
than the results of IllustrisTNG simulation. The predictions of \cite{pillepich18} are systematically higher than the data by \cite{demaio20}, because of a higher slope and similar intercept, and the widest gap is seen 
at high halo masses. Our ICL distribution is less concentrated toward the innermost regions than the simulated one, which translate into a lower amount of ICL in the central 100 kpc. In order to reach this conclusion, we 
developed a detailed analysis by comparing our predictions with IllustrisTNG's results. We showed that both predict the same amount of  BCG+ICL and BCG+ICL+satellite stellar mass within the clusters, and 
different amounts of ICL outside 100 kpc as a function of halo mass, that are separated by $1-\sigma$, with our predictions being higher than those of IllustrisTNG. Considering that our predictions better match the 
observed data, our profile gives a more accurate description of the ICL mass distribution within haloes, which is less concentrated than that invoked by \cite{pillepich18} (their equation 1).  This statement has to be 
taken with caution and needs an explanation. Semi-analytic models, as reminded above, do not provide spatial information on the distribution of any component (while simulations do) and we obtained the distribution of the 
ICL by assuming a NFW like profile with spherical symmetry. However, the fact that haloes of similar mass in Pillepiich et al. have been stacked together with the assumption of spherical symmetry in order to get their functional 
form (their equation 1) makes the two models perfectly comparable in a statistical sense, and the differences in the results are not given by the assumption of spherical symmetry.

Moreover, as mentioned in Section \ref{sec:results}, an important caveat must be discussed and concerns the definitions of the components considered. In fact, \cite{pillepich18} do not address the problem of splitting the 
BCG from the ICL, and call as "central", i.e. BCG+(part of)ICL, all the stellar mass within a given aperture excluding the contribution from satellite galaxies. On the other hand, we are able to distinguish between the 
two components and our profile can perfectly isolate the amount of ICL within any given aperture. Since we assume that the BCG is confined within 100 kpc, the two definitions converge to the same (and so the problem 
drops) if in what they call "central", the BCG stellar mass in their haloes is also confined within 100 kpc, which is likely the case for most of them.

In the last part of our analysis we tested our model against recent observed data by \cite{gonzalez13} and \cite{kravtsov18}, at $z=0$. We plot the amount of BCG and ICL, and that amount with the contribution of 
satellite galaxies within $R_{500}$, as a function of $M_{500}$, and the fraction of stellar mass in the BCG+ICL within $R_{500}$ also as a function of $M_{500}$. In both cases we found a good 
agreement with the available observational data, although an important caveat is worth mentioning and discussing here and concerns the way the amount of BCG+ICL stellar mass has been computed observationally. \cite{gonzalez13} improved the measurements of the former data presented in \cite{gonzalez07} by using a fainter normalization when converting from magnitudes to luminosities that results in having 
luminosities (and so stellar masses) $\sim 15$\% higher. In addition, when converting from luminosities to stellar masses, they also revisited the mass-to-light ratio used in \cite{gonzalez07}. They considered a correction 
from dark matter estimated to be $\sim 15$\%, which results in a mass-to-light ratio about 26\% lower than before, and by considering all corrections they found stellar masses around 13\% lower.  Then, with respect to 
their previous work, the revisited data in \cite{gonzalez13} are closer to our predictions. 

Another important caveat must be mentioned and concerns both the comparison with \cite{gonzalez13} and \cite{kravtsov18}, i.e. their definition of the ICL. \cite{kravtsov18} do not separate the BCG from its outer component
(the ICL) and assume that the outer component can extend up to 200 kpc (270 and 340 kpc in two cases). They also note that in the sample of Gonzalez et al. the ICL is traced up to 300 kpc. This definition is not 
consistent with ours because our assumption is that the ICL extend over all the halo out to $R_{200}$. Although we did not show it, in the range of mass investigated by Gonzalez et al. and Kravtsov et al., according to 
our profile with $\gamma=3$, the amount of ICL outside 300 kpc and within $R_{500}$ can account from $\sim15$\% to $\sim 20$\% of the total stellar mass. This means that, in the right panel of Figure \ref{fig:iclbcg_halom}, 
our predictions are somewhat higher than the observed data also because of the radial cut in measuring the ICL in the two observations.

The idea of describing the stellar mass distribution of the ICL through a modified version of the NFW profile for the dark matter is original, but recently there have been a few suggestions that the ICL could easily trace 
the dark matter distribution.  \cite{montes19} used a sample of six clusters form the Hubble Frontier Fields (\citealt{lotz17}) and compared the bi-dimensional distribution of the dark matter with that of the ICL by using the 
Modified Hausdorff distance (MHD). The MHD is a way of connecting the two distributions and quantifying their similarities and basically gives an idea of how far the two components are from each other (see their paper 
for further details). With that method they found that the average distance between ICL and dark matter is MHD $\sim 25$kpc (within 140 kpc from the center), a result which shows that the ICL follows the global dark matter distribution and can be used 
as a tracer of it.  Similar conclusions have been taken by \cite{kluge20}, who investigated on the ICL-cluster alignment with a sample of around 50 local clusters obtained with the Wendelstein Telescope Wide Field Imager.
In order to qualify the ICL as a good tracer of the dark matter they examined four different criteria, including the ICL-cluster alignment, the BCG-cluster center offset and the ellipticity (the fourth criteria is the line of sight
velocity). They concluded that the ICL is better aligned than the BCG with the host cluster in terms of both position and centering, making the ICL a better tracer of the dark matter than the BCG. Even more recently, 
\cite{alonso20} used the Cluster-EAGLE simulations (\citealt{barnes17,bahe17}) to test the observational result of \cite{montes19} quoted above. They used the same procedure used in Montes et al. and concluded that the 
stellar mass distribution follows that of the total (dark matter included), although their radial profiles differ substantially.

These results are not directly comparable with our predictions shown in Figure \ref{sec:append}\ref{fig:dmicl_diff} and discussed in Appendix \ref{sec:append} since, in general, they differ in terms of method, redshift and halo mass range, but provide the hint that the ICL and the dark matter are really somewhat separated, and quantifying their separation in the next future with more observational data (by confirming the results of \citealt{montes19} even at lower redshift) can be very useful in order to set the paramenter $\gamma$ of our model. Indeed, as already stated above, $\gamma$ can be a function of halo properties such as the mass, or even redshift dependent, and also the only variable in the profile that accounts for the separation between the two components. We will aim to a detailed study on that when more observational data (that we can use to set the profile at different redshift and halo masses) are available.

\section{Conclusions}
\label{sec:conclusions}

We have taken advantage of a semi-analytic model of galaxy formation with a state-of-art implementation of the formation and evolution of the ICL to study its mass distribution along galaxy groups and clusters, 
in a wide range of halo mass and from redshift $z=1.5$ to the present time. Given the fact that the semi-analytic model itself does not provide any spatial distribution, we have introduced a new profile for the ICL 
which is linked to that of the dark matter, motivated by recent observational and theoretical works. We have assumed that the ICL can be described by a simple NFW with a different concentration, which is coupled 
with that of the dark matter halo by the relation $c_{ICL}=\gamma c_{DM}$. By means of observational data of the $M^*_{100}-M_{500}$ relation (where $M^*_{100}$ is the BCG+ICL stellar mass within 100 kpc and 
$M_{500}$ the mass of the halo) at different redshifts, we have set the values of $\gamma$ that best reproduced the observed data considered, and used those values to make a full comparison with other 
observational and theoretical works. From our analysis we can conclude the following:
\begin{itemize}
	\item The fraction of BCG+ICL mass within the innermost 100 kpc is an increasing function of  redshift and $\gamma$, and a decreasing function of the mass of the halo. This can be explained by the fact that 
	         less massive haloes are more concentrated and smaller than more massive ones, and that haloes of the same mass are, on average, less concentrated at higher redshift;
	\item The model is able to reproduce the $M^*_{100}-M_{500}$ relation with $\gamma=3$ at the present time when compared with data by \cite{demaio20}, and closer than the prediction of IllustrisTNG 
	         (\citealt{pillepich18}) simulation. At higher redshift, the model prefers slightly higher values of $\gamma$ and, despite it can reproduce the slope of the relation, especially at $z\sim0.5$ it predicts an intercept that is 
	         lower than that found by DeMaio et al. by around 0.2 dex;
	\item A detailed comparison with the results of IllustrisTNG simulation shows that, although our predictions agrees very well in terms of amount or fraction of BCG+ICL (even including satellite galaxies) mass within 
	         $R_{200}$ (in fairly good agreement with observational data either), the distributions of the ICL along the halo are different. Our profile is less concentrated than that suggested by \cite{pillepich18}, and this explains 
	         why our $M^*_{100}-M_{500}$ relation at the present time matches the observed one, while theirs is higher (similar intercept but a higher slope).
	\item By looking at the distance between the mass distributions of the ICL and the dark matter at $z=0$ (see details in Appendix \ref{sec:append}), we find that low concentrated and more massive haloes have their 
	         dark matter and ICL distributions farther to each other than high concentrated and low mass haloes. 
\end{itemize} 
We suggest that a modified version of the NFW profile with a higher concentration can roughly describe the mass distribution of the ICL in wide ranges of halo mass and redshift. In order to better describe the ICL 
distribution in groups and clusters, more work is needed. The parameter $\gamma$ that we have introduced, and that links the concentration of the ICL to the dark matter one, can be better constrained with the 
help of more observational data in a variety of redshifts. Our aim for a future work is to find the possible dependences of $\gamma$ on halo mass and redshift, i.e a relation such as $\gamma=\gamma(M_{200},z)$.

\section*{Acknowledgements}
The authors thank the anonymous referee for his/her constructive comments which helped to improve the manuscript.
This work is supported by the National Key Research and Development Program of China (No. 2017YFA0402703), 
and by the National Natural Science Foundation of China (Key Project No. 11733002). 
\label{lastpage}

\appendix

\section{ICL and Dark Matter Mass Distributions}
\label{sec:append}

\begin{figure*} 
\begin{center}
\begin{tabular}{cc}
\includegraphics[scale=.49]{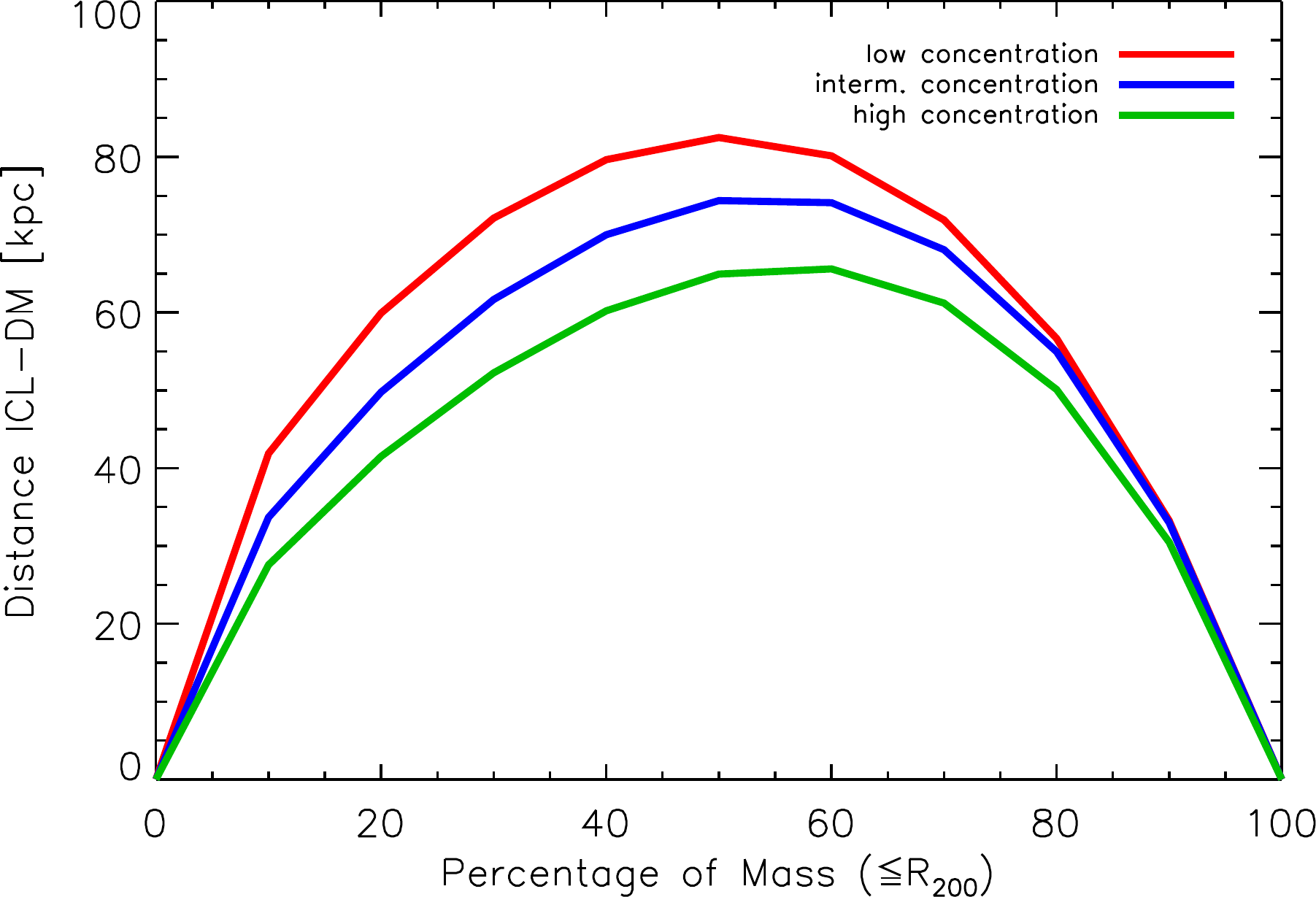} &d
\includegraphics[scale=.49]{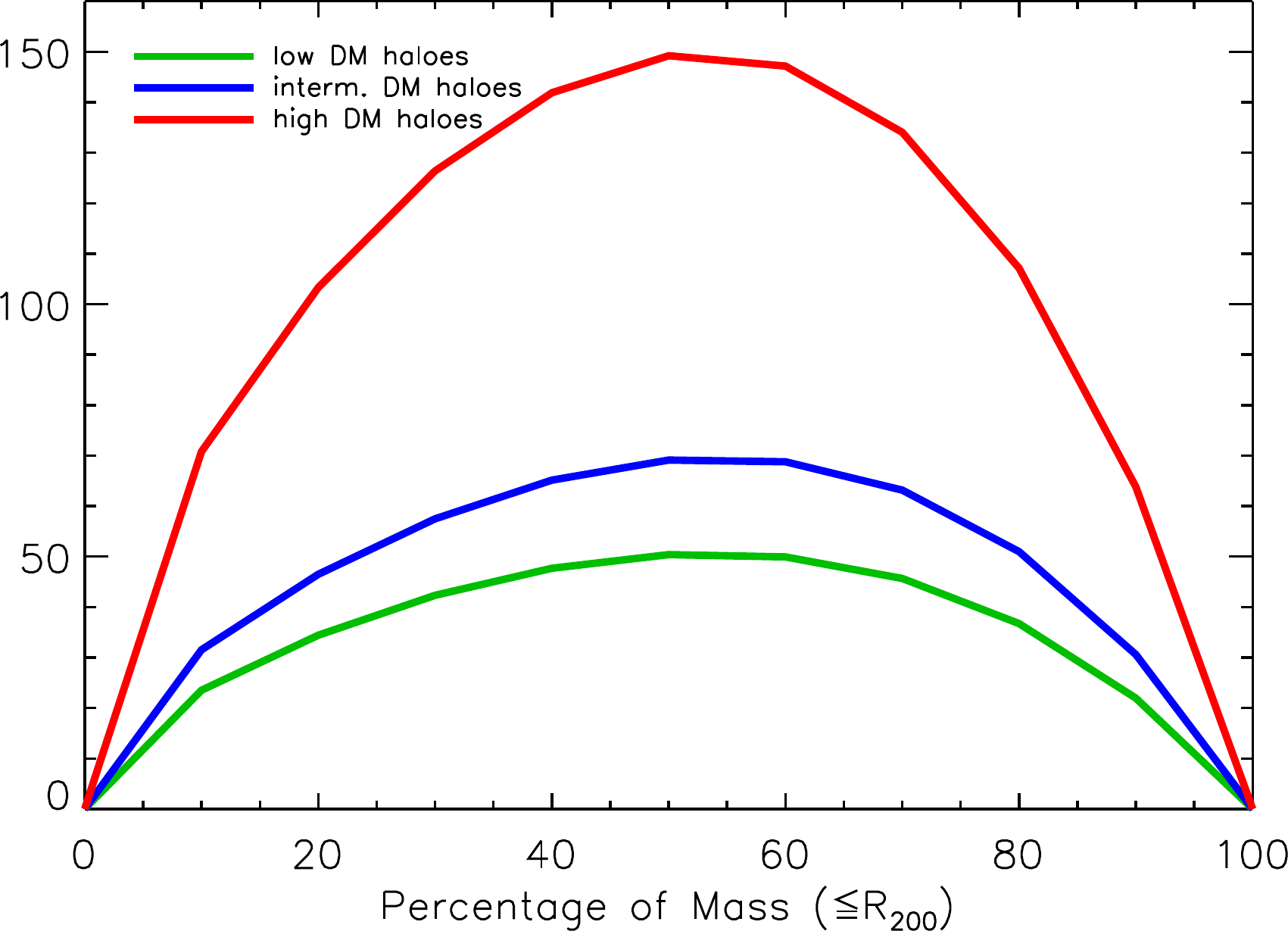} \\
\end{tabular}
\caption{Distance between the mass distributions of dark matter and ICL as a function of the percentage of the total mass that the distributions consider, at $z=0$. In the left panel our sample of haloes are 
divided according to their concentration, while in the right panel they are divided according to the halo mass. In both cases the sample has been split in order to have three subsamples with 
approximately the same number of objects (details are given in the text).  Low concentrated haloes, which also correspond to the more massive, show to have their dark matter and ICL distributions more 
distant with respect to high concentrated and low mass haloes.}
\label{fig:dmicl_diff}
\end{center}
\end{figure*}

We have linked the distribution of the ICL to the distribution of the dark matter through the concentration of the halo \footnote{We remind the reader  that $c_{ICL}=\gamma c_{DM}$.}, so it is natural to wonder how far the ICL and the dark matter mass distributions are from each other, assuming a given value of $\gamma$. We address this point in Figure \ref{sec:append}\ref{fig:dmicl_diff}, where we plot the distance (intended as the separation in kpc between the radii that enclose a given percentage of the mass of the two components, assuming spherical symmetry) between the two mass distributions, ICL and dark matter, as a function of the percentage of the total mass that the distributions consider, at $z=0$.
In the left panel we divide the full sample in three subsamples according to the concentration of the halo, from low ($c_{DM}<4$) to high concentrations ($c_{DM}>5)$, while in the right panel we divide the sample in subsamples 
according to the halo mass, from low mass ($\log M_{200} < 13.5$) to high mass ($\log M_{200} > 14$). In both cases, the three subsamples have been chosen to have roughly the same number of haloes. The picture 
is that low concentrated haloes (red line in the left panel), which also roughly correspond to the most massive (red line in the right panel), have their dark matter and ICL distributions more distant with respect to high concentrated 
(green line in the left panel) and low mass (green line in the right panel) haloes. If we focus on the halo mass, which is a property of the halo that can be easily inferred, there is a remarkable difference in the distance 
of the two mass distributions between low mass haloes and high mass haloes. If we consider half of the mass distributions, in low mass haloes they are around 50 kpc distant, while in high mass haloes they are three 
times more distant, around 150 kpc. These results are qualitatively compared with other works and discussed in Section \ref{sec:discussion}.

\begin{thebibliography}{}

\bibitem[Alonso Asensio et al.(2020)]{alonso20} Alonso Asensio, I., Dalla Vecchia, C., Bah{\'e}, Y.~M., et al.\ 2020, \mnras, 494, 1859
\bibitem[Bah{\'e} et al.(2017)]{bahe17} Bah{\'e}, Y.~M., Barnes, D.~J., Dalla Vecchia, C., et al.\ 2017, \mnras, 470, 4186
\bibitem[Barnes et al.(2017)]{barnes17} Barnes, D.~J., Kay, S.~T., Bah{\'e}, Y.~M., et al.\ 2017, \mnras, 471, 1088
\bibitem[Burke et al.(2015)]{burke15} Burke, C., Hilton, M., \& Collins, C.\ 2015, \mnras, 449, 2353
\bibitem[Chabrier(2003)]{chabrier03} Chabrier, G.\ 2003, \pasp, 115, 763
\bibitem[Contini et al.(2012)]{contini12} Contini, E., De Lucia, G., \& Borgani, S.\ 2012, \mnras, 420, 2978
\bibitem[Contini et al.(2014)]{contini14} Contini, E., De Lucia, G., Villalobos, {\'A}., \& Borgani, S.\ 2014, \mnras, 437, 3787
\bibitem[Contini et al.(2018)]{contini18} Contini, E., Yi, S.~K., \& Kang, X.\ 2018, \mnras, 479, 932
\bibitem[Contini et al.(2019)]{contini19} Contini, E., Yi, S.~K., \& Kang, X.\ 2019, \apj, 871, 24
\bibitem[De Lucia \& Blaizot(2007)]{delucia07} De Lucia, G., \& Blaizot, J.\ 2007, \mnras, 375, 2
\bibitem[DeMaio et al.(2015)]{demaio15} DeMaio, T., Gonzalez, A.~H., Zabludoff, A., Zaritsky, D., \& Brada{\v c}, M.\ 2015, \mnras, 448, 1162
\bibitem[DeMaio et al.(2018)]{demaio18} DeMaio, T., Gonzalez, A.~H., Zabludoff, A., et al.\ 2018, \mnras, 474, 3009
\bibitem[DeMaio et al.(2020)]{demaio20} DeMaio, T., Gonzalez, A.~H., Zabludoff, A., et al.\ 2020, \mnras, 491, 3751
\bibitem[Gao et al.(2004)]{gao04} Gao, L., White, S.~D.~M., Jenkins, A., et al.\ 2004, \mnras, 355, 819
\bibitem[Gao et al.(2011)]{gao11} Gao, L., Frenk, C.~S., Boylan-Kolchin, M., et al.\ 2011, \mnras, 410, 2309
\bibitem[Gonzalez et al.(2007)]{gonzalez07} Gonzalez, A.~H., Zaritsky, D., \& Zabludoff, A.~I.\ 2007, \apj, 666, 147
\bibitem[Gonzalez et al.(2013)]{gonzalez13} Gonzalez, A.~H., Sivanandam, S., Zabludoff, A.~I., et al.\ 2013, \apj, 778, 14
\bibitem[Groenewald et al.(2017)]{groenewald17} Groenewald, D.~N., Skelton, R.~E., Gilbank, D.~G., \& Loubser, S.~I.\ 2017, \mnras, 467, 4101
\bibitem[Han et al.(2018)]{han18} Han, S., Smith, R., Choi, H., et al.\ 2018, \apj, 866, 78
\bibitem[Iodice et al.(2017)]{iodice17} Iodice, E., Spavone, M., Cantiello, M., et al.\ 2017, \apj, 851, 75
\bibitem[Iodice et al.(2020)]{iodice20} Iodice, E., Spavone, M., Cattapan, A., et al.\ 2020, \aap, 635, A3
\bibitem[Kluge et al.(2020)]{kluge20} Kluge, M., Neureiter, B., Riffeser, A., et al.\ 2020, \apjs, 247, 43
\bibitem[Kravtsov et al.(2018)]{kravtsov18} Kravtsov, A.~V., Vikhlinin, A.~A., \& Meshcheryakov, A.~V.\ 2018, Astronomy Letters, 44, 8
\bibitem[Lotz et al.(2017)]{lotz17} Lotz, J.~M., Koekemoer, A., Coe, D., et al.\ 2017, \apj, 837, 97
\bibitem[Montes \& Trujillo(2018)]{montes18} Montes, M., \& Trujillo, I.\ 2018, \mnras, 474, 917
\bibitem[Montes \& Trujillo(2019)]{montes19} Montes, M., \& Trujillo, I.\ 2019, \mnras, 482, 2838
\bibitem[Morishita et al.(2017)]{morishita17} Morishita, T., Abramson, L.~E., Treu, T., et al.\ 2017, \apj, 846, 139
\bibitem[Muldrew et al.(2011)]{muldrew11} Muldrew, S.~I., Pearce, F.~R., \& Power, C.\ 2011, \mnras, 410, 2617
\bibitem[Murante et al.(2007)]{murante07} Murante, G., Giovalli, M., Gerhard, O., et al.\ 2007, \mnras, 377, 2
\bibitem[Navarro et al.(1997)]{navarro97} Navarro, J.~F., Frenk, C.~S., \& White, S.~D.~M.\ 1997, \apj, 490, 493
\bibitem[Pillepich et al.(2018)]{pillepich18} Pillepich, A., Nelson, D., Hernquist, L., et al.\ 2018, \mnras, 475, 648
\bibitem[Prada et al.(2012)]{prada12} Prada, F., Klypin, A.~A., Cuesta, A.~J., et al.\ 2012, \mnras, 423, 3018
\bibitem[Presotto et al.(2014)]{presotto14} Presotto, V., Girardi, M., Nonino, M., et al.\ 2014, \aap, 565, A126
\bibitem[Puchwein et al.(2010)]{puchwein10} Puchwein, E., Springel, V., Sijacki, D., \& Dolag, K.\ 2010, \mnras, 406, 936
\bibitem[Purcell et al.(2007)]{purcell07} Purcell, C.~W., Bullock, J.~S., \& Zentner, A.~R.\ 2007, \apj, 666, 20
\bibitem[Rudick et al.(2011)]{rudick11} Rudick, C.~S., Mihos, J.~C., \& McBride, C.~K.\ 2011, \apj, 732, 48
\bibitem[Springel et al.(2001)]{springel01} Springel, V., White, S.~D.~M., Tormen, G., et al.\ 2001, \mnras, 328, 726
\bibitem[Springel(2010)]{springel10} Springel, V.\ 2010, \mnras, 401, 791
\bibitem[Tang et al.(2018)]{tang18} Tang, L., Lin, W., Cui, W., et al.\ 2018, \apj, 859, 85
\bibitem[Zwicky(1937)]{zwicky37} Zwicky, F.\ 1937, \apj, 86, 217




\end{thebibliography}
\end{document}